\documentclass[
preprint,
superscriptaddress,
reprint,
preprintnumbers,
 amsmath,amssymb,
 aps,
prl
floatfix,
]{revtex4-1}
\usepackage{amsmath}
\usepackage{marvosym}
\usepackage{graphicx,subfigure}         
\usepackage{amsmath} 
\usepackage{amssymb,stmaryrd}   
\usepackage{float}
\usepackage{color}
\usepackage{wrapfig}
\usepackage{multirow}
\usepackage{mathtools}
\usepackage{tikz}  
\usepackage{xspace}
\usetikzlibrary{arrows,shapes,trees,positioning}  
\usepackage{comment} 
\usepackage{bm}
\usepackage{physics}
\usepackage{hyperref}
\usepackage{textcomp}
\hypersetup{
    colorlinks=true,
    linkcolor=red,
    citecolor=red,
    urlcolor=red,
}

\usepackage[T1]{fontenc}
\usepackage{lmodern} 

\usepackage{graphicx}%

\usepackage{gensymb}

\newcommand{\mycomment}[1]{}

\begin{document}

\title{\Large{Anisotropic magnon transport in an antiferromagnetic trilayer heterostructure: is BiFeO$_3$ an altermagnet?}}


\author{Sajid Husain}
\altaffiliation[Authors]{ contributed equally}
\affiliation{Department of Materials Science and Engineering, University of California, Berkeley, CA, USA}
\email{shusain@berkeley.edu}
\affiliation{Department of Materials Science and NanoEngineering, Rice University, Houston, TX, USA}

\author{Maya Ramesh}
\altaffiliation[Authors]{ contributed equally}
\affiliation{Department of Materials Science and Engineering, Cornell University, Ithaca, USA}

\author{Qian Song}
\affiliation{Department of Materials Science and NanoEngineering, Rice University, Houston, TX, USA}

\author{Sergei Prokhorenko}
\affiliation{Smart Ferroic Materials Center, Physics Department and Institute for Nanoscience and Engineering,
University of Arkansas, Fayetteville, Arkansas 72701, USA}

\author{Shashank Kumar Ojha}
\affiliation{Department of Materials Science and NanoEngineering, Rice University, Houston, TX, USA}
\affiliation{Rice Advanced Materials Institute, Rice University, Houston, TX, 77005, USA}

\author{Surya Narayan Panda}
\affiliation{Department of Materials Science and NanoEngineering, Rice University, Houston, TX, USA}
\affiliation{Rice Advanced Materials Institute, Rice University, Houston, TX, 77005, USA}

\author{Xinyan Li}
\affiliation{Rice Advanced Materials Institute, Rice University, Houston, TX, 77005, USA}

\author{Yousra Nahas}
\affiliation{Smart Ferroic Materials Center, Physics Department and Institute for Nanoscience and Engineering,
University of Arkansas, Fayetteville, Arkansas 72701, USA}

\author{Yogesh Kumar}
\affiliation{Department of Materials Science and Engineering, University of California, Berkeley, CA, USA}

\author{Pushpendra Gupta}
\affiliation{Department of Materials Science and Engineering, University of California, Berkeley, CA, USA}

\author{Tenzin Chang}
\affiliation{Department of Materials Science and Engineering, University of California, Berkeley, CA, USA}

\author{Alan Ji-in Jung}
\affiliation{Department of Materials Science and Engineering, University of California, Berkeley, CA, USA}

\author{Rogério de Sousa}
\affiliation{Department of Physics and Astronomy, University of Victoria, Victoria, British Columbia, Canada V8W 2Y2}
\affiliation{Centre for Advanced Materials and Related Technology, University of Victoria, Victoria, British Columbia, Canada V8W 2Y2}

\author{James G. Analytis}
\affiliation{Department of Physics, University of California, Berkeley, CA, USA}

\author{Lane W. Martin}
\affiliation{Department of Materials Science and NanoEngineering, Rice University, Houston, TX, USA}
\affiliation{Rice Advanced Materials Institute, Rice University, Houston, TX, 77005, USA}
\affiliation{Department of Physics and Astronomy, Rice University, Houston, TX, USA}
\affiliation{Department of Chemistry, Rice University, Houston, TX, USA}


\author{Zhi Yao}
\affiliation{Applied Mathematics and Computational Research Division, Lawrence Berkeley National Laboratory, USA, Berkeley, CA 94560, USA}

\author{Sang-Wook Cheong}
\affiliation{Keck Center for Quantum Magnetism and Department of Physics and Astronomy, Rutgers University, Piscataway, New Jersey 08854, USA}

\author{Laurent Bellaiche}
\affiliation{Smart Ferroic Materials Center, Physics Department and Institute for Nanoscience and Engineering,
University of Arkansas, Fayetteville, Arkansas 72701, USA}
\affiliation{Department of Materials Science and Engineering, Tel Aviv University, Ramat Aviv, Tel Aviv 6997801, Israel}

\author{Manuel Bibes}
\affiliation{Laboratoire Albert Fert, CNRS, Thales, Université Paris-Saclay, 91767 Palaiseau, France}

\author{Darrell G. Schlom}
\affiliation{Department of Materials Science and Engineering, Cornell University, Ithaca, USA}
\affiliation{Kavli Institute at Cornell for Nanoscale Science, Cornell University, Ithaca, NY, 14853, USA}
\affiliation{Leibniz-Institut fur Kristallzüchtung, Max-Born-Str. 2, 12489, Berlin, Germany}

\author{Ramamoorthy Ramesh}
\email{rramesh@berkeley.edu}
\affiliation{Department of Materials Science and Engineering, University of California, Berkeley, CA, USA}
\affiliation{Department of Materials Science and NanoEngineering, Rice University, Houston, TX, USA}
\affiliation{Rice Advanced Materials Institute, Rice University, Houston, TX, 77005, USA}
\affiliation{Department of Physics, University of California, Berkeley, CA, USA}
\affiliation{Department of Physics and Astronomy, Rice University, Houston, TX, USA}
\affiliation{Materials Science Division, Lawrence Berkeley National Laboratory, Berkeley, CA, USA}

\begin{abstract}
Magnons provide a route to ultra-fast transport and non-destructive readout of spin-based information transfer. Here, we report magnon transport and its emergent anisotropic nature in BiFeO$_3$ layers confined between ultrathin layers of the antiferromagnet LaFeO$_3$. Due to the confined state, BiFeO$_3$ serves as an efficient magnon transmission channel as well as a magnetoelectric knob by which to control the stack by means of an electric field. We discuss the mechanism of the anisotropic spin transport based on the interaction between the antiferromagnetic order and the electric field. This allows us to manipulate and amplify the spin transport in such a confined geometry. Furthermore, lower crystal symmetric and suppression of the spin cycloid in ultrathin BiFeO$_3$ stabilizes a non-trivial antiferromagnetic state exhibiting symmetry-protected spin-split bands that provide the non-trivial sign inversion of the spin current, which is a characteristic of an altermagnet. This work provides an understanding of the anisotropic spin transport in complex antiferromagnetic heterostructures where ferroelectricity and altermagnetism coexist, paving the way for a new route to realize electric-field control of a novel state of magnetism.
\end{abstract}

\maketitle
Magnons (collective spin excitations) in insulating antiferromagnets (AFMs) offer the potential for ultrafast spin transmission without Joule heating, enabling long-range, non-volatile collective spin transfer \cite{lebrun2018tunable,Parsonnet_Nonvolatile_2022,Huang2024ManipulatingPolarization}. In certain AFMs or two-dimensional ferrimagnets, magnons can transport spin information non-locally over sub-millimeter distances \cite{lebrun2018tunable,Wei2022GiantFilms}; which is several orders of magnitude longer than typical spin diffusion lengths in metals. This enhanced transport arises from collective excitations and the influence of domain structure \cite{fan2023coherent}. Efficient antiferromagnetic magnon transport in insulating AFMs is based on the transfer of angular momentum, mediated by the Néel vector, the canted magnetic order, or both \cite{das2022anisotropic}. Traditionally, magnetic fields have been used to control magnon transport in AFMs \cite{lebrun2018tunable}; however, this method is energetically inefficient, and device output voltages are limited by the properties of spin-orbit-coupled metals. More recently, all-oxide heterostructures have emerged as promising platforms for enhancing output voltages \cite{Huang2024ManipulatingPolarization}, as they enable magnon confinement \cite{Beairsto2021ConfinedMagnons} in AFMs and allow for non-volatile, electric field-based control of magnetism \cite{Husain2025ColossalAntiferromagnet}—a better energy-efficient approach.

\begin{figure*}[t!]
\centering
\includegraphics[width=1.0\textwidth]{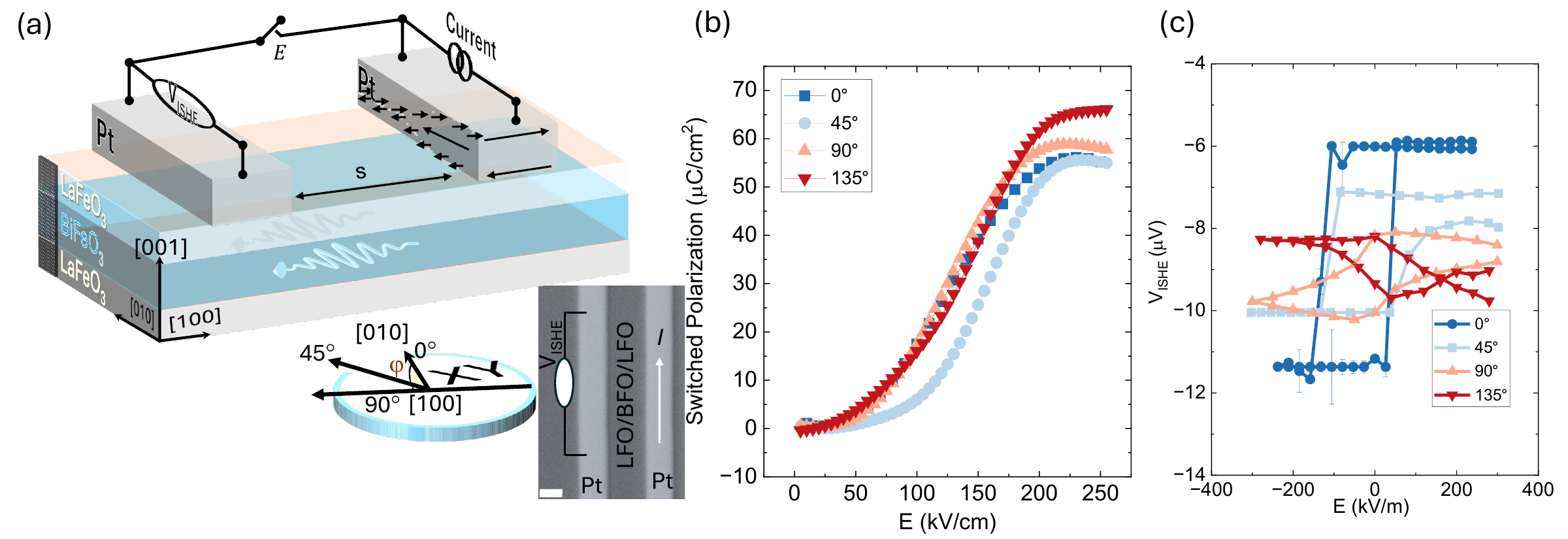}
\caption{(a) Device schematic to measure the non-local inverse spin Hall effect (ISHE) voltage generated by antiferromagnetic magnons excited through the SHE (black arrows on the right Pt electrode). Light shaded, wavy arrow, in the BFO middle layer represents the magnon flow direction. $\varphi$ is the angle between the long axis of the Pt wire and $[010]$, $s$ is the spacing between the metal wires, which is $\sim$1.5 $\mu$m unless otherwise specified. Bottom right is the top-view of the device imaged by the scanning electron microscopy, scale bar is 1$\mu$m. (b) Ferroelectric pulsed polarization measured for various devices under the  PUND scheme. PUND: positive-up-negative-down, is a standard protocol to record the switched ferroelectric polarization under short electrical pulses (4$\mu$s) and eliminates parasitic contributions to the charge response from resistive leakage and dielectric responses. (c) $V_{ISHE}$ hysteresis measured from the devices with various in-plane angles with respect to the substrate, shown in (a). The hysteresis measured at various angles has been centered with respect to the $0\degree$ data to facilitate comparison.}
\label{Fig:Fig.1}
\end{figure*}

In this context, magnetoelectric materials, such as BiFeO$_3$ (BFO), which exhibits robust coupling between antiferromagnetic and ferroelectric order at room temperature, provide a compelling opportunity for electrical control of AFM magnons \cite{liu2021electric}. BFO in its non-cycloidal (non-collinear \cite{cheong2024altermagnetism} or collinear \cite{Bernardini2024Ruddlesden-PopperAltermagnetism}) phase is a promising candidate for hosting a new state of magnetism known as an \textit{altermagnetism}, which requires \( \mathcal{PT} \) and \( \mathcal{T} \)\textit{\textbf{t}} symmetries to be broken. Ferroelectric materials are ideal candidates to break \( \mathcal{PT} \) due to their non-centrosymmetric structure. Moreover, in BFO, octahedral rotation in the $R3c$ phase does break the \textbf{t} (translational) symmetry between the moments of Fe. Thus, it fulfills all the requirements to be classified as an altermagnet \cite{urruPRB_2025}. Furthermore, the spin cycloid needs in BFO to be suppressed, which is achievable via epitaxial strain \cite{bai2005destruction,Dufour2023OnsetFilms,haykal2020antiferromagnetic}, electric field \cite{de2013theory,Ojha2025MorphogenesisAntiferromagnet}, rare-earth substitution \cite{Sajid_LBFO_2024}, high magnetic fields \cite{pervez2025continuous}, or reducing thickness to a few nanometers \cite{Dufour2023OnsetFilms} that satisfies the final criteria for it to become an altermagnet \cite{Sando2013,urruPRB_2025}. Importantly, the intrinsic anisotropy of the electronic band structure of an altermagnet offers a compelling route to probe their magnetic state \cite{gonzalez2023spontaneous,jo2025weak,galindez2025revealing}. Anisotropic spin textures in momentum space can serve as fingerprints of altermagnetic order, especially when combined with angle-resolved charge/spin transport measurements \cite{Smejkal2022b,das2022anisotropic}. 

In this Letter, we report observations of spin transport anisotropy in antiferromagnetic LaFeO$_3$/BiFeO$_3$/LaFeO$_3$ (LFO/BFO/LFO) trilayer heterostructures and the possible existence of an altermagnetic state which can be electrically controlled. We utilize a non-local spin transport (in-plane) device geometry to excite the magnons in the BFO layer via the spin-Hall effect (SHE), and the inverse-spin-Hall effect (ISHE) is used to detect the response as a function of the varying in-plane angles of the non-local devices with respect to the substrate crystallographic directions. The measured anisotropy and the sign reversal in the ISHE voltage provide a signature of an altermagnetic state. 

\begin{figure*}[t!]
\centering
\includegraphics[width=0.9\textwidth]{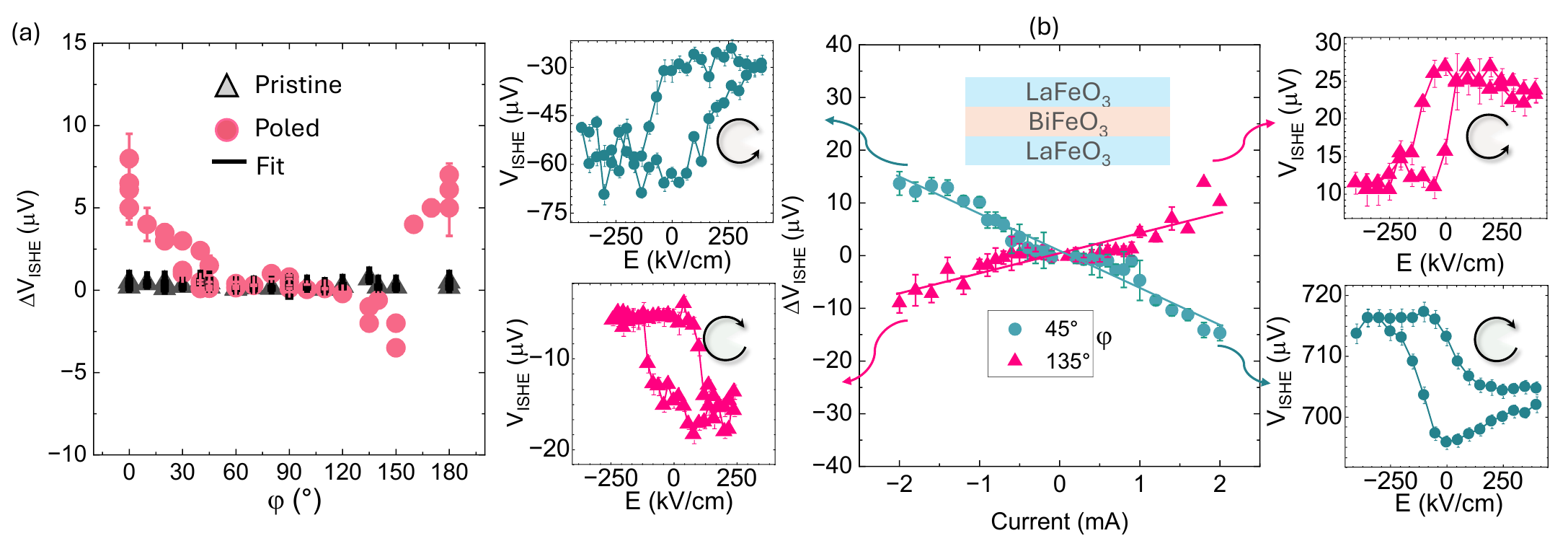}
\caption{The differential voltage ($\Delta V_{ISHE}$) as a function of the angle ($\varphi$) with respect to the substrate edge $[100]$. The Pt wires are parallel to $[010]$ and the electric field is applied along $\varphi$ to measure the polarization and spin transport hysteresis. The data in red have been recorded in the pristine state when the electric field was not applied. The pristine state measurements were simply conducted by swapping the source and drain connections (Supplementary Figure 5~\cite{SM}). (b) $\Delta V_{ISHE}$ as a function of injected current in two devices of $\varphi=45\degree$ and $\varphi=135\degree$ in the LFO/BFO/LFO trilayer. The sign of the $\Delta V_{ISHE}$ gets reversed when the injected current direction is opposite. The hysteresis in the left and right panels is of opposite polarity due to two opposite currents used in the source Pt wire.  Similarly, in the orthogonal devices ($\varphi=45\degree$ or $\varphi=135\degree$), the corresponding hysteresis also shows opposite polarities. The hystereses shown here were measured at a 2mA supply DC.}
\label{Fig:Fig.2}
\end{figure*}

Thin film heterostructures of 5 nm LFO/5 nm BFO/5 nm LFO were deposited on SrTiO$_3$ (001) substrate using molecular-beam epitaxy (Supplementary Information: Methods \cite{SM}). The data presented in the main text was collected on the samples deposited on STO substrate unless otherwise specified. For BFO layers of such thickness, prior work has revealed that the cycloidal magnetic texture is destroyed (Supplementary Information Figure 1 \cite{SM}); this state is expected to be an altermagnet \cite{Bernardini2024Ruddlesden-PopperAltermagnetism}. In the as-grown state, BFO is found to be an antipolar antiferromagnet. An applied electric field converts this anti-polar state into a polar state, accompanied by a $\sim$100-fold increase in the spin transmission, as probed through inverse-spin-Hall measurements \cite{Husain2025ColossalAntiferromagnet}. We hypothesized that this enhancement originates from magnon confinement in the BFO layers due to a differential magnon group velocity \cite{chen2017group} between the orthoferrite LFO ($\sim$14-30 km/s) \cite{das2022anisotropic,Park2018Low-energyBi} and magnetoelectric BFO ($\sim$40-100 km/s) \cite{rovillain2010electric,de2008electrical}. Due to the magnetoelectric nature of the BFO, any changes in its spin texture are picked up in the electric-field dependent non-local spin transport experiments. Such an epitaxial heterostructure forms the model system for us to probe the possibility of an altermagnetic state using the test structure shown in Fig.~\ref{Fig:Fig.1} (a).


Figure~\ref{Fig:Fig.1}(a) depicts the measurement scheme where platinum (Pt) serves as the SHE (source) and ISHE (drain) electrodes. The substrate's in-plane crystallographic axes (Fig.~\ref{Fig:Fig.1}a) were used as a reference frame to fabricate test structures with systematically varying azimuthal angles, $\varphi$, as depicted by the schematic compass in Fig.~\ref{Fig:Fig.1}(a) (Supplementary Figure 2 \cite{SM}). Electric ($E$) field pulses are applied between the two electrodes to switch the polar (magnetic) state of BFO; Supplementary Figure 3(b) shows the comparison of non-local ISHE voltage measured as a function of $E$-field in three representative samples: BFO/Pt, LFO/Pt, and the [LFO/BFO/LFO]/Pt. The LFO/Pt does not show an $E$-field dependence in spin transport (c.f. Supplementary Figure 3(b)~\cite{SM}) due to its non-polar nature, whereas BFO/Pt shows a small spin Seebeck effect as reported previously~\cite{Parsonnet_Nonvolatile_2022,Sajid_LBFO_2024}. A large output voltage in the trilayer is evidence of promising magnon transport, likely due to the magnon confinement \cite{Husain2025ColossalAntiferromagnet}, which motivates us to explore the anisotropy in spin transport. Figure~\ref{Fig:Fig.1}(b) presents an exemplar $E$-field pulsed ferroelectric polarization employing PUND: positive-up-negative-down (see Supplementary Figure 4 \cite{SM}) scheme for 0$\degree$, 45$\degree$, 90$\degree$, and 135$\degree$, illustrating that there is very little difference in the ferroelectric switching behavior as a function of the azimuthal angle, $\varphi$.

An electric field is applied in the $\varphi$ direction to record the data as a function of azimuthal angles. After this field poling step, the spin transport measurements were carried out on the same set of test structures as in Fig.~\ref{Fig:Fig.1}(a) for a range of $\varphi$. For a fixed magnitude of supply current ($I_{ac}$) at the right electrode, the non-local inverse-spin-Hall voltage ($V_{ISHE}$) is recorded (at the left electrode) as a function of $E$-field, examples of which are shown in Fig.~\ref{Fig:Fig.1}(c). The first positive $E$-field pulse is applied along the [100] direction. The peak-to-peak value ($\Delta V_{ISHE}$) changes as a function of $\varphi$, revealing the anisotropic behavior of magnon transport in the heterostructures. Note that the hysteresis polarity reversal occurred at $135\degree$, characterized by a lower peak-to-peak ISHE voltage magnitude. This sign inversion is the key observation of our work, which is the first experimental evidence of an altermagnetic feature emerging in a BFO-based system, which is consistent with other systems \cite{gonzalez2023spontaneous,jo2025weak,galindez2025revealing}. 

Before we address the altermagnetic origin of the anisotropy, Fig. \ref{Fig:Fig.1} (c), we first eliminate the other sources of anisotropy. First, PUND measurements (Figure~\ref{Fig:Fig.1}(b)) rule out a ferroelectric origin of the observed anisotropy. Second, the SQUID magnetometer (Supplementary Figure 2 (c) \cite{SM}) indicates, in-plane magnetic anisotropy is also not expected. We have performed multiple control experiments to rule out other sources of anisotropy (if any), see Supplementary Figure~6, 8, 9~\cite{SM}.

Armed with this, we now discuss the angle dependence of the ISHE voltage, shown in Figure~\ref{Fig:Fig.2}(a) for both the pristine and poled state. The data in the pristine state (in black filled symbol) does not show an angle dependence (see Supplementary Figure 5~\cite{SM} for the measurement scheme). The data in red, Fig.~\ref{Fig:Fig.2}(a), shows the emergence of the anisotropy in the differential ISHE voltage ($\Delta V_{ISHE}$, a difference between the two saturated states, Fig.\ref{Fig:Fig.1}(c)) as a function of in-plane angle in the poled state. The data in Fig.~\ref{Fig:Fig.2}(b) represents the $\Delta V_{ISHE}$ as a function of supply current in the devices patterned at two representative angles, such as $\varphi$ = 45$\degree$ and 135$\degree$. In this panel, positive/negative direct current (Methods, Supplementary Information \cite{SM}) gives $V_{ISHE}$ hysteresis with counter-clockwise/clockwise polarity (represented by the curved arrows in Fig.~\ref{Fig:Fig.2}), which is indicative of the spin Hall effect origin of the magnon excitation at the source electrode. Note that for the positive/negative direct current, there is an additional polarity reversal (pink vs turquoise color), which is a signature of altermagnetism in the heterostructure. We now turn to a microscopic interpretation of this anisotropy and the sign inversion.

\begin{figure}[t!]
\centering
\includegraphics[width=0.5\textwidth]{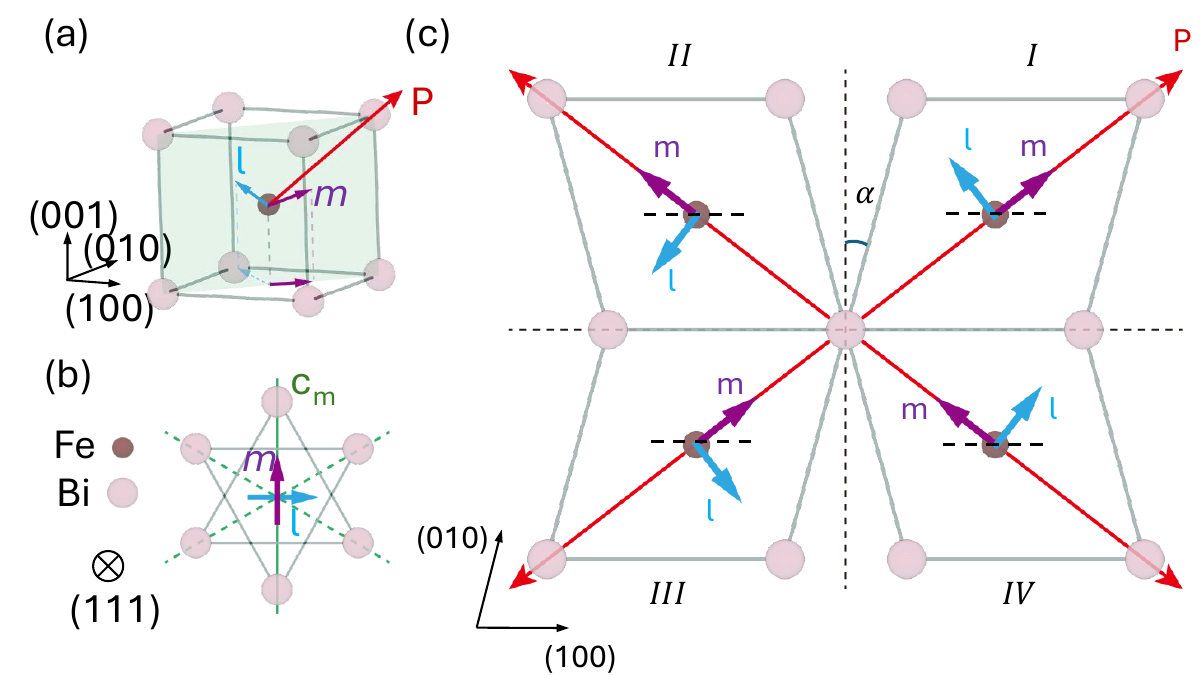}
\caption{\textbf{Description of the anisotropy and sign inversion:}(a) Representation of the rhombohedral \textit{R3c} structure of BFO, illustrating its relationship to the ideal cubic perovskite lattice. The threefold symmetry axis along the pseudocubic [111] direction is marked by a red arrow. (b) A view along the $[111]$ direction. The $c_m$-glide planes are highlighted with dotted black lines. (c) Top view of the four possible situations of the devices at angles of  0$\degree$, 45$\degree$, 90$\degree$, and 135$\degree$ being under consideration under poled conditions. $\alpha$ is the tilt due to the non-centrosymmetric $R3c$, $l$ and $m$ are the antiferromagnetic and canted moment of BFO. $I, II, III$ and $IV$ correspond to the four 4-quadrants that provide access to explain four device angles and connected symmetries.}
\label{Fig:Fig.3}
\end{figure}

Figure~\ref{Fig:Fig.3} represents the schematic of the BFO symmetry at the unit cell level. The polar and magnetic order parameters viewed down the $[111]$ direction with the glide mirror planes ($C_m$) are shown in Figure~\ref{Fig:Fig.3} (a,b). In the device patterned at $\varphi = \sim 0\degree$, reversing the electric field direction from $[1 0 0]_{pc}$ to $[\Bar{1} 0 0]_{pc}$, switches the $\textbf{P}$ in-plane by 71$\degree$ (from $I$-to-$II$, Fig.~\ref{Fig:Fig.3}(c)), which then switches the Néel vector $\textbf{\textit{l}}$ and canted moment ($m$) connected through a rotation. The output voltage at the detector wire exhibits hysteresis, with the magnitude of the voltage being high or low, depending on the net antiferromagnetic and canted moments sensed by the SHE metal wire (See Supplementary Figure 10(a) \cite{SM}). The switching mechanism in the device at $\varphi= \sim 90\degree$ (\textit{i.e.}, along $[100]$), is connected by the mirror plane shown by a horizontal dotted line in Fig.~\ref{Fig:Fig.3}(c), which gives the minimum change ($\Delta V_{ISHE}$) in spin transport (See Supplementary Figure 10(b) \cite{SM}). 

We now apply a similar analysis to the devices at $\sim$45$\degree$ and $\sim$135$\degree$, in order to understand the sign inversion observed in Fig.~\ref{Fig:Fig.2}(b). Switching (polar/magnetic) operations $I-III$ and $II-IV$ (Fig.~\ref{Fig:Fig.3}(c)) produces an equivalent output voltage in the devices at $\sim$45$\degree$ and $\sim$135$\degree$, since these are connected through a mirror operation. However, due to the opposite projection of the net $l$ and/or $m$ around these angles, the output $\Delta V_{ISHE}$ will have opposite sign (See Supplementary Figure 10(c) and (d) \cite{SM}, also see END MATTER for detailed description).

In order to clarify the observed anisotropic spin transport and sign inversion, we also analyzed our data based on the Néel vector relationship between the LFO and BFO, in Supplementary Figure 11 \cite{SM} and in Supplementary Note 1 \cite{SM}. Under non-equilibrium magnon accumulation, $\nabla \left(\mu_1-\mu_2\right)\neq0$ (where $\mu_1$ and $\mu_2$ correspond to the spin accumulation due to the sublattice magnetization $m_1$, and $m_2$ of an antiferromagnet), the generated non-local ISHE voltage,  $V_{ISHE} \propto \cos^{2}\varphi$, which explains the anisotropic spin transport behavior (Fig.~\ref{Fig:Fig.2}(a)). However, this by itself cannot explain the sign inversion. Therefore, the origin of polarity inversion of the ISHE voltage output shown in Figure \ref{Fig:Fig.2} lies in the electronic band structure of BFO, i.e, its altermagnetism.

\begin{figure*}[t!]
\centering
\includegraphics[width=0.85\textwidth]{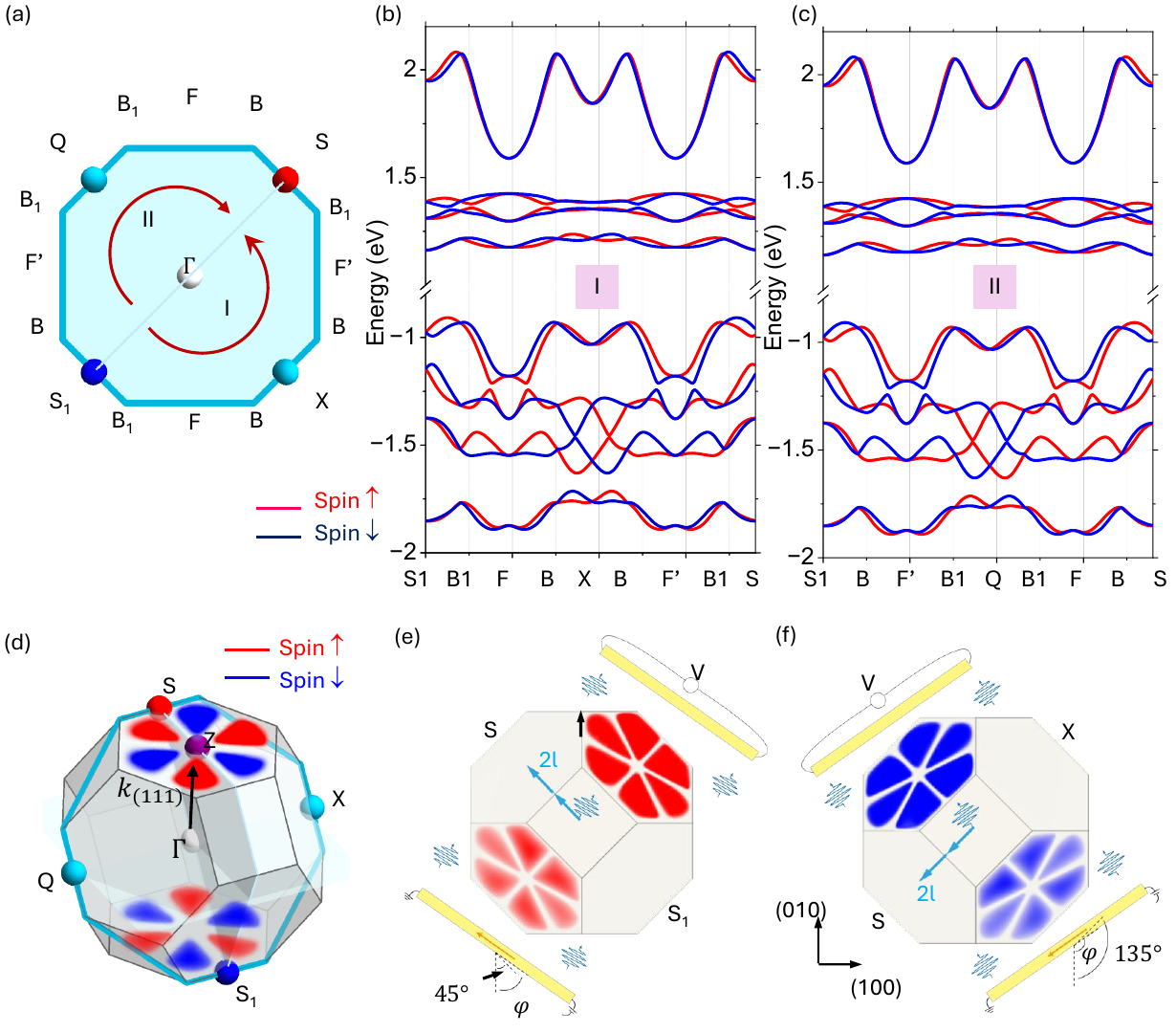}
\caption{\textbf{Anisotropic magnons and emergence of altermagnetism in BiFeO$_3$(BFO)}. (a) Plane view of the $(001)_{\mathrm{pc}}$ Brillouin zone cross-section. Two considered $k$-point paths connecting S and S$_1$ points lie at the boundary of the Brillouin zone and are related via the $(1\bar{1}0)_{\mathrm{pc}}$ mirror plane. Calculated Kohn-Sham band structure of BFO along the two k-point paths shown in panels (b) and (c). Red and blue lines indicate the spin-up and spin-down bands, respectively. (d) First Brillouin zone of the rhombohedral unit cell of $R3c$ bulk BFO with overlapped red and blue spin density bands. Point $X$ corresponds to a $(1\bar{1}0)_{\mathrm{pc}}$ mirror image of $Q$. The corresponding mirror plane containing $P$, $Z$ and $P_1$ points is shown as a gray polygon with light blue edges. Solid blue line passing through $S$, $X$, $S_1$ and $Q$ highlights the boundary of the $(001)_{\mathrm{pc}}$ plane cross-section of the Brillouin zone. (e), (f) The differential spin texture in the band structure corresponds to the devices patterned at $\varphi=$ 45$\degree$ and $\varphi=$ 135$\degree$, respectively. Dark/light color contrast represents the high and low signal, generating the hysteresis between the two polar states.}
\label{Fig:Fig.4}
\end{figure*}

The altermagnetism in BFO is intrinsically related to the alternating oxygen octahedra rotations that surround Fe$^{3+}$ ions, thereby breaking the pure translational symmetry between the spin-up and spin-down AFM sublattices. In the $R3c$ phase of bulk BFO, the two G-AFM sublattices are related by a glide plane that combines translation along the $[111]$ axis and a (1$\bar{1}$0) mirror reflection. As a result, a (1$\bar{1}$0) mirror operation relating $[100]$ to $[010]$ as well as $[1\bar{1}0]$ and [$\bar{1}$10] directions (in pseudocubic notation) is intrinsically linked to an inversion of the spin-up and spin-down channels. This argument is supported by our DFT calculations (END MATTER) presented in Fig.~\ref{Fig:Fig.4}. Figure~\ref{Fig:Fig.4}(a) depicts the Brillouin zone (BZ) of the $R3c$ phase of bulk BFO with the aforementioned $[100]$, $[010]$, $[110]$ and [$\bar{1}$10] zone axes corresponding to the $F'$, $F$, $X$ and $Q$ high symmetry points at the BZ boundary, respectively. The spin-resolved Kohn-Sham bands computed along the two paths at the boundary of the (001) cross-section of the Brillouin zone (Fig.~\ref{Fig:Fig.4} (a)) are shown in Fig.~\ref{Fig:Fig.4} (b,c). The chosen paths are related by a (1$\bar{1}$0) mirror and, as can be seen from Fig.~\ref{Fig:Fig.4} (a), are indeed equivalent upto an inversion of the spin-up and spin-down bands. Such a band splitting is most pronounced within the top valence bands reaching $\sim$0.15 eV in the vicinity of the $F$ and $F'$ points. These calculations support the altermagnetic character of the polar BFO layer. The spin texture drawn (red 'up'/blue 'down') in Fig.~\ref{Fig:Fig.4} (d) determines the sign of the output signal in the spin transport. The schematics of the spin density differential for the devices at $\sim$45$\degree$ (red) and $\sim$135$\degree$ (blue) show the opposite spin contrast (also see END MATTER FIG.EM\ref{Fig:Fig.EM1}) imprinted on the Pt detectors near these angles, which provides a source of opposite polarity observed in the spin transport when BFO is an altermagnet.

In conclusion, we demonstrate a strong anisotropy in spin transport in an antiferromagnetic trilayer heterostructure in which BFO serves as a source of magnon confinement that can be controlled by an electric field. We also observed the sign reversal in the charge-spin interconversion, which is indicative of an altermagnetic origin from BFO. 
The intrinsic magnetoelectric coupling allows electric fields to manipulate the altermagnetic phase, directly tying the existence and control of altermagnetic order to the ferroelectric state. Thus, non-cycloidal BFO stands as a prototypical material where the interplay of altermagnetism and ferroelectricity is possible.

This work was supported by the U.S. Department of Energy, Office of Science, Office of Basic Energy Sciences, Materials Sciences and Engineering Division under Contract No. DE-AC02-05CH11231 within the Quantum Materials Program (No. KC2202). S.P., L.B., L.W.M., D.G.S., M.R., and R.R. acknowledge support from the Army Research Office under the ETHOS MURI via cooperative agreement W911NF-21-2-0162 and the Army Research Laboratory under Cooperative Agreement Number W911NF-24-2-0100. The views and conclusions contained in this document are those of the authors and should not be interpreted as representing the official policies, either expressed or implied, of the Army Research Laboratory or the U.S. Government. The U.S. Government is authorized to reproduce and distribute reprints for Government purposes, notwithstanding any copyright notation herein. S.P. and L.B. additionally acknowledge support from the Vannevar Bush Faculty Fellowship (VBFF) from the Department of Defense and an Impact Grant 3.0 from ARA. Y.H., L.W.M., and R.R. also acknowledge the support of the National Science Foundation under Grant DMR-2329111. We thank Tenzin Chang and Alan Ji-in Jung for helping with the codes.

\bibliographystyle{apsrev4-1}
\bibliography{aps.bib}

\providecommand{\noopsort}[1]{}\providecommand{\singleletter}[1]{#1}%
\begin{thebibliography}{29}%
\makeatletter
\providecommand \@ifxundefined [1]{%
 \@ifx{#1\undefined}
}%
\providecommand \@ifnum [1]{%
 \ifnum #1\expandafter \@firstoftwo
 \else \expandafter \@secondoftwo
 \fi
}%
\providecommand \@ifx [1]{%
 \ifx #1\expandafter \@firstoftwo
 \else \expandafter \@secondoftwo
 \fi
}%
\providecommand \natexlab [1]{#1}%
\providecommand \enquote  [1]{``#1''}%
\providecommand \bibnamefont  [1]{#1}%
\providecommand \bibfnamefont [1]{#1}%
\providecommand \citenamefont [1]{#1}%
\providecommand \href@noop [0]{\@secondoftwo}%
\providecommand \href [0]{\begingroup \@sanitize@url \@href}%
\providecommand \@href[1]{\@@startlink{#1}\@@href}%
\providecommand \@@href[1]{\endgroup#1\@@endlink}%
\providecommand \@sanitize@url [0]{\catcode `\\12\catcode `\$12\catcode `\&12\catcode `\#12\catcode `\^12\catcode `\_12\catcode `\%12\relax}%
\providecommand \@@startlink[1]{}%
\providecommand \@@endlink[0]{}%
\providecommand \url  [0]{\begingroup\@sanitize@url \@url }%
\providecommand \@url [1]{\endgroup\@href {#1}{\urlprefix }}%
\providecommand \urlprefix  [0]{URL }%
\providecommand \Eprint [0]{\href }%
\providecommand \doibase [0]{http://dx.doi.org/}%
\providecommand \selectlanguage [0]{\@gobble}%
\providecommand \bibinfo  [0]{\@secondoftwo}%
\providecommand \bibfield  [0]{\@secondoftwo}%
\providecommand \translation [1]{[#1]}%
\providecommand \BibitemOpen [0]{}%
\providecommand \bibitemStop [0]{}%
\providecommand \bibitemNoStop [0]{.\EOS\space}%
\providecommand \EOS [0]{\spacefactor3000\relax}%
\providecommand \BibitemShut  [1]{\csname bibitem#1\endcsname}%
\let\auto@bib@innerbib\@empty
\bibitem [{\citenamefont {Lebrun}\ \emph {et~al.}(2018)\citenamefont {Lebrun}, \citenamefont {Ross}, \citenamefont {Bender}, \citenamefont {Qaiumzadeh}, \citenamefont {Baldrati}, \citenamefont {Cramer}, \citenamefont {Brataas}, \citenamefont {Duine},\ and\ \citenamefont {Kl{\"a}ui}}]{lebrun2018tunable}%
  \BibitemOpen
  \bibfield  {author} {\bibinfo {author} {\bibfnamefont {R.}~\bibnamefont {Lebrun}}, \bibinfo {author} {\bibfnamefont {A.}~\bibnamefont {Ross}}, \bibinfo {author} {\bibfnamefont {S.}~\bibnamefont {Bender}}, \bibinfo {author} {\bibfnamefont {A.}~\bibnamefont {Qaiumzadeh}}, \bibinfo {author} {\bibfnamefont {L.}~\bibnamefont {Baldrati}}, \bibinfo {author} {\bibfnamefont {J.}~\bibnamefont {Cramer}}, \bibinfo {author} {\bibfnamefont {A.}~\bibnamefont {Brataas}}, \bibinfo {author} {\bibfnamefont {R.}~\bibnamefont {Duine}}, \ and\ \bibinfo {author} {\bibfnamefont {M.}~\bibnamefont {Kl{\"a}ui}},\ }\href@noop {} {\bibfield  {journal} {\bibinfo  {journal} {Nature}\ }\textbf {\bibinfo {volume} {561}},\ \bibinfo {pages} {222} (\bibinfo {year} {2018})}\BibitemShut {NoStop}%
\bibitem [{\citenamefont {Parsonnet}\ \emph {et~al.}(2022)\citenamefont {Parsonnet}, \citenamefont {Caretta}, \citenamefont {Nagarajan}, \citenamefont {Zhang}, \citenamefont {Taghinejad}, \citenamefont {Behera}, \citenamefont {Huang}, \citenamefont {Kavle}, \citenamefont {Fernandez}, \citenamefont {Nikonov}, \citenamefont {Li}, \citenamefont {Young}, \citenamefont {Analytis},\ and\ \citenamefont {Ramesh}}]{Parsonnet_Nonvolatile_2022}%
  \BibitemOpen
  \bibfield  {author} {\bibinfo {author} {\bibfnamefont {E.}~\bibnamefont {Parsonnet}}, \bibinfo {author} {\bibfnamefont {L.}~\bibnamefont {Caretta}}, \bibinfo {author} {\bibfnamefont {V.}~\bibnamefont {Nagarajan}}, \bibinfo {author} {\bibfnamefont {H.}~\bibnamefont {Zhang}}, \bibinfo {author} {\bibfnamefont {H.}~\bibnamefont {Taghinejad}}, \bibinfo {author} {\bibfnamefont {P.}~\bibnamefont {Behera}}, \bibinfo {author} {\bibfnamefont {X.}~\bibnamefont {Huang}}, \bibinfo {author} {\bibfnamefont {P.}~\bibnamefont {Kavle}}, \bibinfo {author} {\bibfnamefont {A.}~\bibnamefont {Fernandez}}, \bibinfo {author} {\bibfnamefont {D.}~\bibnamefont {Nikonov}}, \bibinfo {author} {\bibfnamefont {H.}~\bibnamefont {Li}}, \bibinfo {author} {\bibfnamefont {I.}~\bibnamefont {Young}}, \bibinfo {author} {\bibfnamefont {J.}~\bibnamefont {Analytis}}, \ and\ \bibinfo {author} {\bibfnamefont {R.}~\bibnamefont {Ramesh}},\ }\href {\doibase 10.1103/PhysRevLett.129.087601} {\bibfield  {journal} {\bibinfo  {journal} {Phys. Rev. Lett.}\
  }\textbf {\bibinfo {volume} {129}},\ \bibinfo {pages} {087601} (\bibinfo {year} {2022})}\BibitemShut {NoStop}%
\bibitem [{\citenamefont {Huang}\ \emph {et~al.}(2024)\citenamefont {Huang}, \citenamefont {Chen}, \citenamefont {Li}, \citenamefont {Mangeri}, \citenamefont {Zhang}, \citenamefont {Ramesh}, \citenamefont {Taghinejad}, \citenamefont {Meisenheimer}, \citenamefont {Caretta}, \citenamefont {Susarla}, \citenamefont {Jain}, \citenamefont {Klewe}, \citenamefont {Wang}, \citenamefont {Chen}, \citenamefont {Hsu}, \citenamefont {Harris}, \citenamefont {Husain}, \citenamefont {Pan}, \citenamefont {Yin}, \citenamefont {Shafer}, \citenamefont {Qiu}, \citenamefont {Rodrigues}, \citenamefont {Heinonen}, \citenamefont {Vasudevan}, \citenamefont {{\'{I}}{\~{n}}iguez}, \citenamefont {Schlom}, \citenamefont {Salahuddin}, \citenamefont {Martin}, \citenamefont {Analytis}, \citenamefont {Ralph}, \citenamefont {Cheng}, \citenamefont {Yao},\ and\ \citenamefont {Ramesh}}]{Huang2024ManipulatingPolarization}%
  \BibitemOpen
  \bibfield  {author} {\bibinfo {author} {\bibfnamefont {X.}~\bibnamefont {Huang}}, \bibinfo {author} {\bibfnamefont {X.}~\bibnamefont {Chen}}, \bibinfo {author} {\bibfnamefont {Y.}~\bibnamefont {Li}}, \bibinfo {author} {\bibfnamefont {J.}~\bibnamefont {Mangeri}}, \bibinfo {author} {\bibfnamefont {H.}~\bibnamefont {Zhang}}, \bibinfo {author} {\bibfnamefont {M.}~\bibnamefont {Ramesh}}, \bibinfo {author} {\bibfnamefont {H.}~\bibnamefont {Taghinejad}}, \bibinfo {author} {\bibfnamefont {P.}~\bibnamefont {Meisenheimer}}, \bibinfo {author} {\bibfnamefont {L.}~\bibnamefont {Caretta}}, \bibinfo {author} {\bibfnamefont {S.}~\bibnamefont {Susarla}}, \bibinfo {author} {\bibfnamefont {R.}~\bibnamefont {Jain}}, \bibinfo {author} {\bibfnamefont {C.}~\bibnamefont {Klewe}}, \bibinfo {author} {\bibfnamefont {T.}~\bibnamefont {Wang}}, \bibinfo {author} {\bibfnamefont {R.}~\bibnamefont {Chen}}, \bibinfo {author} {\bibfnamefont {C.~H.}\ \bibnamefont {Hsu}}, \bibinfo {author} {\bibfnamefont {I.}~\bibnamefont {Harris}}, \bibinfo
  {author} {\bibfnamefont {S.}~\bibnamefont {Husain}}, \bibinfo {author} {\bibfnamefont {H.}~\bibnamefont {Pan}}, \bibinfo {author} {\bibfnamefont {J.}~\bibnamefont {Yin}}, \bibinfo {author} {\bibfnamefont {P.}~\bibnamefont {Shafer}}, \bibinfo {author} {\bibfnamefont {Z.}~\bibnamefont {Qiu}}, \bibinfo {author} {\bibfnamefont {D.~R.}\ \bibnamefont {Rodrigues}}, \bibinfo {author} {\bibfnamefont {O.}~\bibnamefont {Heinonen}}, \bibinfo {author} {\bibfnamefont {D.}~\bibnamefont {Vasudevan}}, \bibinfo {author} {\bibfnamefont {J.}~\bibnamefont {{\'{I}}{\~{n}}iguez}}, \bibinfo {author} {\bibfnamefont {D.~G.}\ \bibnamefont {Schlom}}, \bibinfo {author} {\bibfnamefont {S.}~\bibnamefont {Salahuddin}}, \bibinfo {author} {\bibfnamefont {L.~W.}\ \bibnamefont {Martin}}, \bibinfo {author} {\bibfnamefont {J.~G.}\ \bibnamefont {Analytis}}, \bibinfo {author} {\bibfnamefont {D.~C.}\ \bibnamefont {Ralph}}, \bibinfo {author} {\bibfnamefont {R.}~\bibnamefont {Cheng}}, \bibinfo {author} {\bibfnamefont {Z.}~\bibnamefont {Yao}}, \ and\
  \bibinfo {author} {\bibfnamefont {R.}~\bibnamefont {Ramesh}},\ }\href@noop {} {\bibfield  {journal} {\bibinfo  {journal} {Nature Materials}\ }\textbf {\bibinfo {volume} {23}},\ \bibinfo {pages} {898} (\bibinfo {year} {2024})}\BibitemShut {NoStop}%
\bibitem [{\citenamefont {Wei}\ \emph {et~al.}(2022)\citenamefont {Wei}, \citenamefont {Santos}, \citenamefont {Lusero}, \citenamefont {Bauer}, \citenamefont {Ben~Youssef},\ and\ \citenamefont {van Wees}}]{Wei2022GiantFilms}%
  \BibitemOpen
  \bibfield  {author} {\bibinfo {author} {\bibfnamefont {X.~Y.}\ \bibnamefont {Wei}}, \bibinfo {author} {\bibfnamefont {O.~A.}\ \bibnamefont {Santos}}, \bibinfo {author} {\bibfnamefont {C.~H.}\ \bibnamefont {Lusero}}, \bibinfo {author} {\bibfnamefont {G.~E.}\ \bibnamefont {Bauer}}, \bibinfo {author} {\bibfnamefont {J.}~\bibnamefont {Ben~Youssef}}, \ and\ \bibinfo {author} {\bibfnamefont {B.~J.}\ \bibnamefont {van Wees}},\ }\href {\doibase 10.1038/s41563-022-01369-0} {\bibfield  {journal} {\bibinfo  {journal} {Nature Materials}\ }\textbf {\bibinfo {volume} {21}},\ \bibinfo {pages} {1352} (\bibinfo {year} {2022})}\BibitemShut {NoStop}%
\bibitem [{\citenamefont {Fan}\ \emph {et~al.}(2023)\citenamefont {Fan}, \citenamefont {Gross}, \citenamefont {Fakhrul}, \citenamefont {Finley}, \citenamefont {Hou}, \citenamefont {Ngo}, \citenamefont {Liu},\ and\ \citenamefont {Ross}}]{fan2023coherent}%
  \BibitemOpen
  \bibfield  {author} {\bibinfo {author} {\bibfnamefont {Y.}~\bibnamefont {Fan}}, \bibinfo {author} {\bibfnamefont {M.~J.}\ \bibnamefont {Gross}}, \bibinfo {author} {\bibfnamefont {T.}~\bibnamefont {Fakhrul}}, \bibinfo {author} {\bibfnamefont {J.}~\bibnamefont {Finley}}, \bibinfo {author} {\bibfnamefont {J.~T.}\ \bibnamefont {Hou}}, \bibinfo {author} {\bibfnamefont {S.}~\bibnamefont {Ngo}}, \bibinfo {author} {\bibfnamefont {L.}~\bibnamefont {Liu}}, \ and\ \bibinfo {author} {\bibfnamefont {C.~A.}\ \bibnamefont {Ross}},\ }\href@noop {} {\bibfield  {journal} {\bibinfo  {journal} {Nature Nanotechnology}\ }\textbf {\bibinfo {volume} {18}},\ \bibinfo {pages} {1000} (\bibinfo {year} {2023})}\BibitemShut {NoStop}%
\bibitem [{\citenamefont {Das}\ \emph {et~al.}(2022)\citenamefont {Das}, \citenamefont {Ross}, \citenamefont {Ma}, \citenamefont {Becker}, \citenamefont {Schmitt}, \citenamefont {van Duijn}, \citenamefont {Galindez-Ruales}, \citenamefont {Fuhrmann}, \citenamefont {Syskaki}, \citenamefont {Ebels} \emph {et~al.}}]{das2022anisotropic}%
  \BibitemOpen
  \bibfield  {author} {\bibinfo {author} {\bibfnamefont {S.}~\bibnamefont {Das}}, \bibinfo {author} {\bibfnamefont {A.}~\bibnamefont {Ross}}, \bibinfo {author} {\bibfnamefont {X.}~\bibnamefont {Ma}}, \bibinfo {author} {\bibfnamefont {S.}~\bibnamefont {Becker}}, \bibinfo {author} {\bibfnamefont {C.}~\bibnamefont {Schmitt}}, \bibinfo {author} {\bibfnamefont {F.}~\bibnamefont {van Duijn}}, \bibinfo {author} {\bibfnamefont {E.~F.}\ \bibnamefont {Galindez-Ruales}}, \bibinfo {author} {\bibfnamefont {F.}~\bibnamefont {Fuhrmann}}, \bibinfo {author} {\bibfnamefont {M.-A.}\ \bibnamefont {Syskaki}}, \bibinfo {author} {\bibfnamefont {U.}~\bibnamefont {Ebels}},  \emph {et~al.},\ }\href@noop {} {\bibfield  {journal} {\bibinfo  {journal} {Nature Communications}\ }\textbf {\bibinfo {volume} {13}},\ \bibinfo {pages} {6140} (\bibinfo {year} {2022})}\BibitemShut {NoStop}%
\bibitem [{\citenamefont {Beairsto}\ \emph {et~al.}(2021)\citenamefont {Beairsto}, \citenamefont {Cazayous}, \citenamefont {Fishman},\ and\ \citenamefont {De~Sousa}}]{Beairsto2021ConfinedMagnons}%
  \BibitemOpen
  \bibfield  {author} {\bibinfo {author} {\bibfnamefont {S.}~\bibnamefont {Beairsto}}, \bibinfo {author} {\bibfnamefont {M.}~\bibnamefont {Cazayous}}, \bibinfo {author} {\bibfnamefont {R.~S.}\ \bibnamefont {Fishman}}, \ and\ \bibinfo {author} {\bibfnamefont {R.}~\bibnamefont {De~Sousa}},\ }\href@noop {} {\bibfield  {journal} {\bibinfo  {journal} {Physical Review B}\ }\textbf {\bibinfo {volume} {104}} (\bibinfo {year} {2021})}\BibitemShut {NoStop}%
\bibitem [{\citenamefont {Husain}\ \emph {et~al.}(2025)\citenamefont {Husain}, \citenamefont {Ramesh}, \citenamefont {Li}, \citenamefont {Prokhorenko}, \citenamefont {Ojha}, \citenamefont {Ross}, \citenamefont {Das}, \citenamefont {Zhao}, \citenamefont {Park}, \citenamefont {Meisenheimer}, \citenamefont {Nahas}, \citenamefont {Caretta}, \citenamefont {Martin}, \citenamefont {Kim}, \citenamefont {Yao}, \citenamefont {Wen}, \citenamefont {Salahuddin}, \citenamefont {Chen}, \citenamefont {Han}, \citenamefont {de~Sousa}, \citenamefont {Bellaiche}, \citenamefont {Bibes}, \citenamefont {Schlom},\ and\ \citenamefont {Ramesh}}]{Husain2025ColossalAntiferromagnet}%
  \BibitemOpen
  \bibfield  {author} {\bibinfo {author} {\bibfnamefont {S.}~\bibnamefont {Husain}}, \bibinfo {author} {\bibfnamefont {M.}~\bibnamefont {Ramesh}}, \bibinfo {author} {\bibfnamefont {X.}~\bibnamefont {Li}}, \bibinfo {author} {\bibfnamefont {S.}~\bibnamefont {Prokhorenko}}, \bibinfo {author} {\bibfnamefont {S.~K.}\ \bibnamefont {Ojha}}, \bibinfo {author} {\bibfnamefont {A.}~\bibnamefont {Ross}}, \bibinfo {author} {\bibfnamefont {K.}~\bibnamefont {Das}}, \bibinfo {author} {\bibfnamefont {B.}~\bibnamefont {Zhao}}, \bibinfo {author} {\bibfnamefont {H.~W.}\ \bibnamefont {Park}}, \bibinfo {author} {\bibfnamefont {P.}~\bibnamefont {Meisenheimer}}, \bibinfo {author} {\bibfnamefont {Y.}~\bibnamefont {Nahas}}, \bibinfo {author} {\bibfnamefont {L.}~\bibnamefont {Caretta}}, \bibinfo {author} {\bibfnamefont {L.~W.}\ \bibnamefont {Martin}}, \bibinfo {author} {\bibfnamefont {S.~K.}\ \bibnamefont {Kim}}, \bibinfo {author} {\bibfnamefont {Z.}~\bibnamefont {Yao}}, \bibinfo {author} {\bibfnamefont {H.}~\bibnamefont {Wen}},
  \bibinfo {author} {\bibfnamefont {S.}~\bibnamefont {Salahuddin}}, \bibinfo {author} {\bibfnamefont {L.-Q.}\ \bibnamefont {Chen}}, \bibinfo {author} {\bibfnamefont {Y.}~\bibnamefont {Han}}, \bibinfo {author} {\bibfnamefont {R.}~\bibnamefont {de~Sousa}}, \bibinfo {author} {\bibfnamefont {L.}~\bibnamefont {Bellaiche}}, \bibinfo {author} {\bibfnamefont {M.}~\bibnamefont {Bibes}}, \bibinfo {author} {\bibfnamefont {D.~G.}\ \bibnamefont {Schlom}}, \ and\ \bibinfo {author} {\bibfnamefont {R.}~\bibnamefont {Ramesh}},\ }\href {https://arxiv.org/abs/2503.23724v1} {\ ,\ \bibinfo {pages} {arxivId: 2503.23724} (\bibinfo {year} {2025})}\BibitemShut {NoStop}%
\bibitem [{\citenamefont {Liu}\ \emph {et~al.}(2021)\citenamefont {Liu}, \citenamefont {Luo}, \citenamefont {Hong}, \citenamefont {Zhang}, \citenamefont {Saglam}, \citenamefont {Li}, \citenamefont {Lin}, \citenamefont {Fisher}, \citenamefont {Pearson}, \citenamefont {Jiang} \emph {et~al.}}]{liu2021electric}%
  \BibitemOpen
  \bibfield  {author} {\bibinfo {author} {\bibfnamefont {C.}~\bibnamefont {Liu}}, \bibinfo {author} {\bibfnamefont {Y.}~\bibnamefont {Luo}}, \bibinfo {author} {\bibfnamefont {D.}~\bibnamefont {Hong}}, \bibinfo {author} {\bibfnamefont {S.~S.-L.}\ \bibnamefont {Zhang}}, \bibinfo {author} {\bibfnamefont {H.}~\bibnamefont {Saglam}}, \bibinfo {author} {\bibfnamefont {Y.}~\bibnamefont {Li}}, \bibinfo {author} {\bibfnamefont {Y.}~\bibnamefont {Lin}}, \bibinfo {author} {\bibfnamefont {B.}~\bibnamefont {Fisher}}, \bibinfo {author} {\bibfnamefont {J.~E.}\ \bibnamefont {Pearson}}, \bibinfo {author} {\bibfnamefont {J.~S.}\ \bibnamefont {Jiang}},  \emph {et~al.},\ }\href@noop {} {\bibfield  {journal} {\bibinfo  {journal} {Science Advances}\ }\textbf {\bibinfo {volume} {7}},\ \bibinfo {pages} {eabg1669} (\bibinfo {year} {2021})}\BibitemShut {NoStop}%
\bibitem [{\citenamefont {Cheong}\ and\ \citenamefont {Huang}(2024)}]{cheong2024altermagnetism}%
  \BibitemOpen
  \bibfield  {author} {\bibinfo {author} {\bibfnamefont {S.-W.}\ \bibnamefont {Cheong}}\ and\ \bibinfo {author} {\bibfnamefont {F.-T.}\ \bibnamefont {Huang}},\ }\href@noop {} {\bibfield  {journal} {\bibinfo  {journal} {npj Quantum Materials}\ }\textbf {\bibinfo {volume} {9}},\ \bibinfo {pages} {13} (\bibinfo {year} {2024})}\BibitemShut {NoStop}%
\bibitem [{\citenamefont {Bernardini}\ \emph {et~al.}(2024)\citenamefont {Bernardini}, \citenamefont {Fiebig},\ and\ \citenamefont {Cano}}]{Bernardini2024Ruddlesden-PopperAltermagnetism}%
  \BibitemOpen
  \bibfield  {author} {\bibinfo {author} {\bibfnamefont {F.}~\bibnamefont {Bernardini}}, \bibinfo {author} {\bibfnamefont {M.}~\bibnamefont {Fiebig}}, \ and\ \bibinfo {author} {\bibfnamefont {A.}~\bibnamefont {Cano}},\ }\href {\doibase 10.1063/5.0252836/3339285} {\bibfield  {journal} {\bibinfo  {journal} {Journal of Applied Physics}\ }\textbf {\bibinfo {volume} {137}},\ \bibinfo {pages} {103903} (\bibinfo {year} {2024})}\BibitemShut {NoStop}%
\bibitem [{\citenamefont {Urru}\ \emph {et~al.}(2025)\citenamefont {Urru}, \citenamefont {Seleznev}, \citenamefont {Teng}, \citenamefont {Park}, \citenamefont {Reyes-Lillo},\ and\ \citenamefont {Rabe}}]{urruPRB_2025}%
  \BibitemOpen
  \bibfield  {author} {\bibinfo {author} {\bibfnamefont {A.}~\bibnamefont {Urru}}, \bibinfo {author} {\bibfnamefont {D.}~\bibnamefont {Seleznev}}, \bibinfo {author} {\bibfnamefont {Y.}~\bibnamefont {Teng}}, \bibinfo {author} {\bibfnamefont {S.~Y.}\ \bibnamefont {Park}}, \bibinfo {author} {\bibfnamefont {S.~E.}\ \bibnamefont {Reyes-Lillo}}, \ and\ \bibinfo {author} {\bibfnamefont {K.~M.}\ \bibnamefont {Rabe}},\ }\href {\doibase 10.1103/v3fg-6smc} {\bibfield  {journal} {\bibinfo  {journal} {Phys. Rev. B}\ }\textbf {\bibinfo {volume} {112}},\ \bibinfo {pages} {104411} (\bibinfo {year} {2025})}\BibitemShut {NoStop}%
\bibitem [{\citenamefont {Bai}\ \emph {et~al.}(2005)\citenamefont {Bai}, \citenamefont {Wang}, \citenamefont {Wuttig}, \citenamefont {Li}, \citenamefont {Wang}, \citenamefont {Pyatakov}, \citenamefont {Zvezdin}, \citenamefont {Cross},\ and\ \citenamefont {Viehland}}]{bai2005destruction}%
  \BibitemOpen
  \bibfield  {author} {\bibinfo {author} {\bibfnamefont {F.}~\bibnamefont {Bai}}, \bibinfo {author} {\bibfnamefont {J.}~\bibnamefont {Wang}}, \bibinfo {author} {\bibfnamefont {M.}~\bibnamefont {Wuttig}}, \bibinfo {author} {\bibfnamefont {J.}~\bibnamefont {Li}}, \bibinfo {author} {\bibfnamefont {N.}~\bibnamefont {Wang}}, \bibinfo {author} {\bibfnamefont {A.~P.}\ \bibnamefont {Pyatakov}}, \bibinfo {author} {\bibfnamefont {A.~K.}\ \bibnamefont {Zvezdin}}, \bibinfo {author} {\bibfnamefont {L.}~\bibnamefont {Cross}}, \ and\ \bibinfo {author} {\bibfnamefont {D.}~\bibnamefont {Viehland}},\ }\href@noop {} {\bibfield  {journal} {\bibinfo  {journal} {Applied Physics Letters}\ }\textbf {\bibinfo {volume} {86}} (\bibinfo {year} {2005})}\BibitemShut {NoStop}%
\bibitem [{\citenamefont {Dufour}\ \emph {et~al.}(2023)\citenamefont {Dufour}, \citenamefont {Abdelsamie}, \citenamefont {Fischer}, \citenamefont {Finco}, \citenamefont {Haykal}, \citenamefont {Sarott}, \citenamefont {Varotto}, \citenamefont {Carr{\'{e}}t{\'{e}}ro}, \citenamefont {Collin}, \citenamefont {Godel}, \citenamefont {Jaouen}, \citenamefont {Viret}, \citenamefont {Trassin}, \citenamefont {Bouzehouane}, \citenamefont {Jacques}, \citenamefont {Chauleau}, \citenamefont {Fusil},\ and\ \citenamefont {Garcia}}]{Dufour2023OnsetFilms}%
  \BibitemOpen
  \bibfield  {author} {\bibinfo {author} {\bibfnamefont {P.}~\bibnamefont {Dufour}}, \bibinfo {author} {\bibfnamefont {A.}~\bibnamefont {Abdelsamie}}, \bibinfo {author} {\bibfnamefont {J.}~\bibnamefont {Fischer}}, \bibinfo {author} {\bibfnamefont {A.}~\bibnamefont {Finco}}, \bibinfo {author} {\bibfnamefont {A.}~\bibnamefont {Haykal}}, \bibinfo {author} {\bibfnamefont {M.~F.}\ \bibnamefont {Sarott}}, \bibinfo {author} {\bibfnamefont {S.}~\bibnamefont {Varotto}}, \bibinfo {author} {\bibfnamefont {C.}~\bibnamefont {Carr{\'{e}}t{\'{e}}ro}}, \bibinfo {author} {\bibfnamefont {S.}~\bibnamefont {Collin}}, \bibinfo {author} {\bibfnamefont {F.}~\bibnamefont {Godel}}, \bibinfo {author} {\bibfnamefont {N.}~\bibnamefont {Jaouen}}, \bibinfo {author} {\bibfnamefont {M.}~\bibnamefont {Viret}}, \bibinfo {author} {\bibfnamefont {M.}~\bibnamefont {Trassin}}, \bibinfo {author} {\bibfnamefont {K.}~\bibnamefont {Bouzehouane}}, \bibinfo {author} {\bibfnamefont {V.}~\bibnamefont {Jacques}}, \bibinfo {author} {\bibfnamefont {J.~Y.}\
  \bibnamefont {Chauleau}}, \bibinfo {author} {\bibfnamefont {S.}~\bibnamefont {Fusil}}, \ and\ \bibinfo {author} {\bibfnamefont {V.}~\bibnamefont {Garcia}},\ }\href {\doibase 10.1021/ACS.NANOLETT.3C02875/SUPPL{\_}FILE/NL3C02875{\_}SI{\_}001.PDF} {\bibfield  {journal} {\bibinfo  {journal} {Nano Letters}\ }\textbf {\bibinfo {volume} {23}},\ \bibinfo {pages} {9073} (\bibinfo {year} {2023})}\BibitemShut {NoStop}%
\bibitem [{\citenamefont {Haykal}\ \emph {et~al.}(2020)\citenamefont {Haykal}, \citenamefont {Fischer}, \citenamefont {Akhtar}, \citenamefont {Chauleau}, \citenamefont {Sando}, \citenamefont {Finco}, \citenamefont {Godel}, \citenamefont {Birkh{\"o}lzer}, \citenamefont {Carr{\'e}t{\'e}ro}, \citenamefont {Jaouen} \emph {et~al.}}]{haykal2020antiferromagnetic}%
  \BibitemOpen
  \bibfield  {author} {\bibinfo {author} {\bibfnamefont {A.}~\bibnamefont {Haykal}}, \bibinfo {author} {\bibfnamefont {J.}~\bibnamefont {Fischer}}, \bibinfo {author} {\bibfnamefont {W.}~\bibnamefont {Akhtar}}, \bibinfo {author} {\bibfnamefont {J.-Y.}\ \bibnamefont {Chauleau}}, \bibinfo {author} {\bibfnamefont {D.}~\bibnamefont {Sando}}, \bibinfo {author} {\bibfnamefont {A.}~\bibnamefont {Finco}}, \bibinfo {author} {\bibfnamefont {F.}~\bibnamefont {Godel}}, \bibinfo {author} {\bibfnamefont {Y.}~\bibnamefont {Birkh{\"o}lzer}}, \bibinfo {author} {\bibfnamefont {C.}~\bibnamefont {Carr{\'e}t{\'e}ro}}, \bibinfo {author} {\bibfnamefont {N.}~\bibnamefont {Jaouen}},  \emph {et~al.},\ }\href@noop {} {\bibfield  {journal} {\bibinfo  {journal} {Nature communications}\ }\textbf {\bibinfo {volume} {11}},\ \bibinfo {pages} {1704} (\bibinfo {year} {2020})}\BibitemShut {NoStop}%
\bibitem [{\citenamefont {de~Sousa}\ \emph {et~al.}(2013)\citenamefont {de~Sousa}, \citenamefont {Allen},\ and\ \citenamefont {Cazayous}}]{de2013theory}%
  \BibitemOpen
  \bibfield  {author} {\bibinfo {author} {\bibfnamefont {R.}~\bibnamefont {de~Sousa}}, \bibinfo {author} {\bibfnamefont {M.}~\bibnamefont {Allen}}, \ and\ \bibinfo {author} {\bibfnamefont {M.}~\bibnamefont {Cazayous}},\ }\href@noop {} {\bibfield  {journal} {\bibinfo  {journal} {Physical Review Letters}\ }\textbf {\bibinfo {volume} {110}},\ \bibinfo {pages} {267202} (\bibinfo {year} {2013})}\BibitemShut {NoStop}%
\bibitem [{\citenamefont {Ojha}\ \emph {et~al.}(2025)\citenamefont {Ojha}, \citenamefont {Pal}, \citenamefont {Prokhorenko}, \citenamefont {Husain}, \citenamefont {Ramesh}, \citenamefont {Li}, \citenamefont {Kang}, \citenamefont {Meisenheimer}, \citenamefont {Schlom}, \citenamefont {Stevenson}, \citenamefont {Caretta}, \citenamefont {Nahas}, \citenamefont {Han}, \citenamefont {Martin}, \citenamefont {Bellaiche}, \citenamefont {Eom},\ and\ \citenamefont {Ramesh}}]{Ojha2025MorphogenesisAntiferromagnet}%
  \BibitemOpen
  \bibfield  {author} {\bibinfo {author} {\bibfnamefont {S.~K.}\ \bibnamefont {Ojha}}, \bibinfo {author} {\bibfnamefont {P.}~\bibnamefont {Pal}}, \bibinfo {author} {\bibfnamefont {S.}~\bibnamefont {Prokhorenko}}, \bibinfo {author} {\bibfnamefont {S.}~\bibnamefont {Husain}}, \bibinfo {author} {\bibfnamefont {M.}~\bibnamefont {Ramesh}}, \bibinfo {author} {\bibfnamefont {X.}~\bibnamefont {Li}}, \bibinfo {author} {\bibfnamefont {D.}~\bibnamefont {Kang}}, \bibinfo {author} {\bibfnamefont {P.}~\bibnamefont {Meisenheimer}}, \bibinfo {author} {\bibfnamefont {D.~G.}\ \bibnamefont {Schlom}}, \bibinfo {author} {\bibfnamefont {P.}~\bibnamefont {Stevenson}}, \bibinfo {author} {\bibfnamefont {L.}~\bibnamefont {Caretta}}, \bibinfo {author} {\bibfnamefont {Y.}~\bibnamefont {Nahas}}, \bibinfo {author} {\bibfnamefont {Y.}~\bibnamefont {Han}}, \bibinfo {author} {\bibfnamefont {L.~W.}\ \bibnamefont {Martin}}, \bibinfo {author} {\bibfnamefont {L.}~\bibnamefont {Bellaiche}}, \bibinfo {author} {\bibfnamefont {C.-B.}\ \bibnamefont
  {Eom}}, \ and\ \bibinfo {author} {\bibfnamefont {R.}~\bibnamefont {Ramesh}},\ }\href {\doibase 10.1073/PNAS.2423298122} {\bibfield  {journal} {\bibinfo  {journal} {Proceedings of the National Academy of Sciences}\ }\textbf {\bibinfo {volume} {122}},\ \bibinfo {pages} {e2423298122} (\bibinfo {year} {2025})}\BibitemShut {NoStop}%
\bibitem [{\citenamefont {Husain}\ \emph {et~al.}(2024)\citenamefont {Husain}, \citenamefont {Harris}, \citenamefont {Meisenheimer}, \citenamefont {Mantri}, \citenamefont {Li}, \citenamefont {Ramesh}, \citenamefont {Behera}, \citenamefont {Taghinejad}, \citenamefont {Kim}, \citenamefont {Kavle} \emph {et~al.}}]{Sajid_LBFO_2024}%
  \BibitemOpen
  \bibfield  {author} {\bibinfo {author} {\bibfnamefont {S.}~\bibnamefont {Husain}}, \bibinfo {author} {\bibfnamefont {I.}~\bibnamefont {Harris}}, \bibinfo {author} {\bibfnamefont {P.}~\bibnamefont {Meisenheimer}}, \bibinfo {author} {\bibfnamefont {S.}~\bibnamefont {Mantri}}, \bibinfo {author} {\bibfnamefont {X.}~\bibnamefont {Li}}, \bibinfo {author} {\bibfnamefont {M.}~\bibnamefont {Ramesh}}, \bibinfo {author} {\bibfnamefont {P.}~\bibnamefont {Behera}}, \bibinfo {author} {\bibfnamefont {H.}~\bibnamefont {Taghinejad}}, \bibinfo {author} {\bibfnamefont {J.}~\bibnamefont {Kim}}, \bibinfo {author} {\bibfnamefont {P.}~\bibnamefont {Kavle}},  \emph {et~al.},\ }\href@noop {} {\bibfield  {journal} {\bibinfo  {journal} {Nature communications}\ }\textbf {\bibinfo {volume} {15}},\ \bibinfo {pages} {5966} (\bibinfo {year} {2024})}\BibitemShut {NoStop}%
\bibitem [{\citenamefont {Pervez}\ \emph {et~al.}(2025)\citenamefont {Pervez}, \citenamefont {Zhang}, \citenamefont {Huang}, \citenamefont {Caretta}, \citenamefont {Ramesh},\ and\ \citenamefont {Ulrich}}]{pervez2025continuous}%
  \BibitemOpen
  \bibfield  {author} {\bibinfo {author} {\bibfnamefont {M.~F.}\ \bibnamefont {Pervez}}, \bibinfo {author} {\bibfnamefont {H.}~\bibnamefont {Zhang}}, \bibinfo {author} {\bibfnamefont {Y.-L.}\ \bibnamefont {Huang}}, \bibinfo {author} {\bibfnamefont {L.}~\bibnamefont {Caretta}}, \bibinfo {author} {\bibfnamefont {R.}~\bibnamefont {Ramesh}}, \ and\ \bibinfo {author} {\bibfnamefont {C.}~\bibnamefont {Ulrich}},\ }\href@noop {} {\bibfield  {journal} {\bibinfo  {journal} {Physical Review B}\ }\textbf {\bibinfo {volume} {111}},\ \bibinfo {pages} {174426} (\bibinfo {year} {2025})}\BibitemShut {NoStop}%
\bibitem [{\citenamefont {Sando}\ \emph {et~al.}(2013)\citenamefont {Sando}, \citenamefont {Barth{\'e}l{\'e}my},\ and\ \citenamefont {Bibes}}]{Sando2013}%
  \BibitemOpen
  \bibfield  {author} {\bibinfo {author} {\bibfnamefont {D.}~\bibnamefont {Sando}}, \bibinfo {author} {\bibfnamefont {A.}~\bibnamefont {Barth{\'e}l{\'e}my}}, \ and\ \bibinfo {author} {\bibfnamefont {M.}~\bibnamefont {Bibes}},\ }\href@noop {} {\bibfield  {journal} {\bibinfo  {journal} {Nature Materials}\ }\textbf {\bibinfo {volume} {12}},\ \bibinfo {pages} {641} (\bibinfo {year} {2013})}\BibitemShut {NoStop}%
\bibitem [{\citenamefont {Gonzalez~Betancourt}\ \emph {et~al.}(2023)\citenamefont {Gonzalez~Betancourt}, \citenamefont {Zub{\'a}{\v{c}}}, \citenamefont {Gonzalez-Hernandez}, \citenamefont {Geishendorf}, \citenamefont {{\v{S}}ob{\'a}{\v{n}}}, \citenamefont {Springholz}, \citenamefont {Olejn{\'\i}k}, \citenamefont {{\v{S}}mejkal}, \citenamefont {Sinova}, \citenamefont {Jungwirth} \emph {et~al.}}]{gonzalez2023spontaneous}%
  \BibitemOpen
  \bibfield  {author} {\bibinfo {author} {\bibfnamefont {R.}~\bibnamefont {Gonzalez~Betancourt}}, \bibinfo {author} {\bibfnamefont {J.}~\bibnamefont {Zub{\'a}{\v{c}}}}, \bibinfo {author} {\bibfnamefont {R.}~\bibnamefont {Gonzalez-Hernandez}}, \bibinfo {author} {\bibfnamefont {K.}~\bibnamefont {Geishendorf}}, \bibinfo {author} {\bibfnamefont {Z.}~\bibnamefont {{\v{S}}ob{\'a}{\v{n}}}}, \bibinfo {author} {\bibfnamefont {G.}~\bibnamefont {Springholz}}, \bibinfo {author} {\bibfnamefont {K.}~\bibnamefont {Olejn{\'\i}k}}, \bibinfo {author} {\bibfnamefont {L.}~\bibnamefont {{\v{S}}mejkal}}, \bibinfo {author} {\bibfnamefont {J.}~\bibnamefont {Sinova}}, \bibinfo {author} {\bibfnamefont {T.}~\bibnamefont {Jungwirth}},  \emph {et~al.},\ }\href@noop {} {\bibfield  {journal} {\bibinfo  {journal} {Physical Review Letters}\ }\textbf {\bibinfo {volume} {130}},\ \bibinfo {pages} {036702} (\bibinfo {year} {2023})}\BibitemShut {NoStop}%
\bibitem [{\citenamefont {Jo}\ \emph {et~al.}(2025)\citenamefont {Jo}, \citenamefont {Go}, \citenamefont {Mokrousov}, \citenamefont {Oppeneer}, \citenamefont {Cheong},\ and\ \citenamefont {Lee}}]{jo2025weak}%
  \BibitemOpen
  \bibfield  {author} {\bibinfo {author} {\bibfnamefont {D.}~\bibnamefont {Jo}}, \bibinfo {author} {\bibfnamefont {D.}~\bibnamefont {Go}}, \bibinfo {author} {\bibfnamefont {Y.}~\bibnamefont {Mokrousov}}, \bibinfo {author} {\bibfnamefont {P.~M.}\ \bibnamefont {Oppeneer}}, \bibinfo {author} {\bibfnamefont {S.-W.}\ \bibnamefont {Cheong}}, \ and\ \bibinfo {author} {\bibfnamefont {H.-W.}\ \bibnamefont {Lee}},\ }\href@noop {} {\bibfield  {journal} {\bibinfo  {journal} {Physical review letters}\ }\textbf {\bibinfo {volume} {134}},\ \bibinfo {pages} {196703} (\bibinfo {year} {2025})}\BibitemShut {NoStop}%
\bibitem [{\citenamefont {Galindez-Ruales}\ \emph {et~al.}(2025)\citenamefont {Galindez-Ruales}, \citenamefont {Gonzalez-Hernandez}, \citenamefont {Schmitt}, \citenamefont {Das}, \citenamefont {Fuhrmann}, \citenamefont {Ross}, \citenamefont {Golias}, \citenamefont {Akashdeep}, \citenamefont {L{\"u}nenb{\"u}rger}, \citenamefont {Baek} \emph {et~al.}}]{galindez2025revealing}%
  \BibitemOpen
  \bibfield  {author} {\bibinfo {author} {\bibfnamefont {E.}~\bibnamefont {Galindez-Ruales}}, \bibinfo {author} {\bibfnamefont {R.}~\bibnamefont {Gonzalez-Hernandez}}, \bibinfo {author} {\bibfnamefont {C.}~\bibnamefont {Schmitt}}, \bibinfo {author} {\bibfnamefont {S.}~\bibnamefont {Das}}, \bibinfo {author} {\bibfnamefont {F.}~\bibnamefont {Fuhrmann}}, \bibinfo {author} {\bibfnamefont {A.}~\bibnamefont {Ross}}, \bibinfo {author} {\bibfnamefont {E.}~\bibnamefont {Golias}}, \bibinfo {author} {\bibfnamefont {A.}~\bibnamefont {Akashdeep}}, \bibinfo {author} {\bibfnamefont {L.}~\bibnamefont {L{\"u}nenb{\"u}rger}}, \bibinfo {author} {\bibfnamefont {E.}~\bibnamefont {Baek}},  \emph {et~al.},\ }\href@noop {} {\bibfield  {journal} {\bibinfo  {journal} {Advanced Materials}\ ,\ \bibinfo {pages} {e05019}} (\bibinfo {year} {2025})}\BibitemShut {NoStop}%
\bibitem [{\citenamefont {Fedchenko}\ \emph {et~al.}(2024)\citenamefont {Fedchenko}, \citenamefont {Min{\'a}r}, \citenamefont {Akashdeep}, \citenamefont {D’Souza}, \citenamefont {Vasilyev}, \citenamefont {Tkach}, \citenamefont {Odenbreit}, \citenamefont {Nguyen}, \citenamefont {Kutnyakhov}, \citenamefont {Wind} \emph {et~al.}}]{Smejkal2022b}%
  \BibitemOpen
  \bibfield  {author} {\bibinfo {author} {\bibfnamefont {O.}~\bibnamefont {Fedchenko}}, \bibinfo {author} {\bibfnamefont {J.}~\bibnamefont {Min{\'a}r}}, \bibinfo {author} {\bibfnamefont {A.}~\bibnamefont {Akashdeep}}, \bibinfo {author} {\bibfnamefont {S.~W.}\ \bibnamefont {D’Souza}}, \bibinfo {author} {\bibfnamefont {D.}~\bibnamefont {Vasilyev}}, \bibinfo {author} {\bibfnamefont {O.}~\bibnamefont {Tkach}}, \bibinfo {author} {\bibfnamefont {L.}~\bibnamefont {Odenbreit}}, \bibinfo {author} {\bibfnamefont {Q.}~\bibnamefont {Nguyen}}, \bibinfo {author} {\bibfnamefont {D.}~\bibnamefont {Kutnyakhov}}, \bibinfo {author} {\bibfnamefont {N.}~\bibnamefont {Wind}},  \emph {et~al.},\ }\href@noop {} {\bibfield  {journal} {\bibinfo  {journal} {Science advances}\ }\textbf {\bibinfo {volume} {10}},\ \bibinfo {pages} {eadj4883} (\bibinfo {year} {2024})}\BibitemShut {NoStop}%
\bibitem [{SM()}]{SM}%
  \BibitemOpen
  \href@noop {} {}\bibinfo {note} {See Supplemental Material at [link] for additional information. Device resistance, magnon transport and NV magnetometry imaging on various control samples. Role of symmetry and emergence of altermagnetism BiFeO$_3$. Analytical modeling of anisotropic magnon transport.}\BibitemShut {Stop}%
\bibitem [{\citenamefont {Chen}\ \emph {et~al.}(2017)\citenamefont {Chen}, \citenamefont {Zakeri}, \citenamefont {Ernst}, \citenamefont {Qin}, \citenamefont {Meng},\ and\ \citenamefont {Kirschner}}]{chen2017group}%
  \BibitemOpen
  \bibfield  {author} {\bibinfo {author} {\bibfnamefont {Y.-J.}\ \bibnamefont {Chen}}, \bibinfo {author} {\bibfnamefont {K.}~\bibnamefont {Zakeri}}, \bibinfo {author} {\bibfnamefont {A.}~\bibnamefont {Ernst}}, \bibinfo {author} {\bibfnamefont {H.}~\bibnamefont {Qin}}, \bibinfo {author} {\bibfnamefont {Y.}~\bibnamefont {Meng}}, \ and\ \bibinfo {author} {\bibfnamefont {J.}~\bibnamefont {Kirschner}},\ }\href@noop {} {\bibfield  {journal} {\bibinfo  {journal} {Physical Review Letters}\ }\textbf {\bibinfo {volume} {119}},\ \bibinfo {pages} {267201} (\bibinfo {year} {2017})}\BibitemShut {NoStop}%
\bibitem [{\citenamefont {Park}\ \emph {et~al.}(2018)\citenamefont {Park}, \citenamefont {Sim}, \citenamefont {Leiner}, \citenamefont {Yoshida}, \citenamefont {Jeong}, \citenamefont {Yano}, \citenamefont {Gardner}, \citenamefont {Bourges}, \citenamefont {Klicpera}, \citenamefont {Sechovsk{\'{y}}}, \citenamefont {Boehm},\ and\ \citenamefont {Park}}]{Park2018Low-energyBi}%
  \BibitemOpen
  \bibfield  {author} {\bibinfo {author} {\bibfnamefont {K.}~\bibnamefont {Park}}, \bibinfo {author} {\bibfnamefont {H.}~\bibnamefont {Sim}}, \bibinfo {author} {\bibfnamefont {J.~C.}\ \bibnamefont {Leiner}}, \bibinfo {author} {\bibfnamefont {Y.}~\bibnamefont {Yoshida}}, \bibinfo {author} {\bibfnamefont {J.}~\bibnamefont {Jeong}}, \bibinfo {author} {\bibfnamefont {S.~I.}\ \bibnamefont {Yano}}, \bibinfo {author} {\bibfnamefont {J.}~\bibnamefont {Gardner}}, \bibinfo {author} {\bibfnamefont {P.}~\bibnamefont {Bourges}}, \bibinfo {author} {\bibfnamefont {M.}~\bibnamefont {Klicpera}}, \bibinfo {author} {\bibfnamefont {V.}~\bibnamefont {Sechovsk{\'{y}}}}, \bibinfo {author} {\bibfnamefont {M.}~\bibnamefont {Boehm}}, \ and\ \bibinfo {author} {\bibfnamefont {J.~G.}\ \bibnamefont {Park}},\ }\href {\doibase 10.1088/1361-648X/aac06b} {\bibfield  {journal} {\bibinfo  {journal} {Journal of Physics Condensed Matter}\ }\textbf {\bibinfo {volume} {30}} (\bibinfo {year} {2018}),\ 10.1088/1361-648X/aac06b}\BibitemShut {NoStop}%
\bibitem [{\citenamefont {Rovillain}\ \emph {et~al.}(2010)\citenamefont {Rovillain}, \citenamefont {De~Sousa}, \citenamefont {Gallais}, \citenamefont {Sacuto}, \citenamefont {M{\'e}asson}, \citenamefont {Colson}, \citenamefont {Forget}, \citenamefont {Bibes}, \citenamefont {Barth{\'e}l{\'e}my},\ and\ \citenamefont {Cazayous}}]{rovillain2010electric}%
  \BibitemOpen
  \bibfield  {author} {\bibinfo {author} {\bibfnamefont {P.}~\bibnamefont {Rovillain}}, \bibinfo {author} {\bibfnamefont {R.}~\bibnamefont {De~Sousa}}, \bibinfo {author} {\bibfnamefont {Y.}~\bibnamefont {Gallais}}, \bibinfo {author} {\bibfnamefont {A.}~\bibnamefont {Sacuto}}, \bibinfo {author} {\bibfnamefont {M.}~\bibnamefont {M{\'e}asson}}, \bibinfo {author} {\bibfnamefont {D.}~\bibnamefont {Colson}}, \bibinfo {author} {\bibfnamefont {A.}~\bibnamefont {Forget}}, \bibinfo {author} {\bibfnamefont {M.}~\bibnamefont {Bibes}}, \bibinfo {author} {\bibfnamefont {A.}~\bibnamefont {Barth{\'e}l{\'e}my}}, \ and\ \bibinfo {author} {\bibfnamefont {M.}~\bibnamefont {Cazayous}},\ }\href@noop {} {\bibfield  {journal} {\bibinfo  {journal} {Nature materials}\ }\textbf {\bibinfo {volume} {9}},\ \bibinfo {pages} {975} (\bibinfo {year} {2010})}\BibitemShut {NoStop}%
\bibitem [{\citenamefont {de~Sousa}\ and\ \citenamefont {Moore}(2008)}]{de2008electrical}%
  \BibitemOpen
  \bibfield  {author} {\bibinfo {author} {\bibfnamefont {R.}~\bibnamefont {de~Sousa}}\ and\ \bibinfo {author} {\bibfnamefont {J.~E.}\ \bibnamefont {Moore}},\ }\href@noop {} {\bibfield  {journal} {\bibinfo  {journal} {Applied Physics Letters}\ }\textbf {\bibinfo {volume} {92}} (\bibinfo {year} {2008})}\BibitemShut {NoStop}%
\end{thebibliography}%

\begin{center}
    \textbf{END MATTER}
\end{center}

\textit{Density Functional Theory:} Density functional theory calculations of the $R3c$ phase of bulk BiFeO$_3$ were performed using Abinit. We employ the Projector Augmented-Wave (PAW) method with PBE (Perdew-Birke-Ernzerhof) GGA approximation for the exchange correlation functional and U correction on Fe d orbitals of 2 eV. The collinear magnetism is treated at the scalar wavefunctions level. We use a 6 $\times$ 6 $\times$ 6 $k$-point mesh for Brillouin zone integrations and a plane-wave cut-off of 25 Ha while the cutoff for the double grid is set to 50 Ha. Self-consistent ground-state calculations are performed with a convergence threshold of 10$^{-10}$ Ha. The relaxation of atomic positions is performed until reaching the tolerance for differences of forces of 10$^{-6}$ hartree/Bohr under the assumption of the fixed unit cell geometry given by the rhombohedral lattice constant of 10.639 Bohr and rhombohedral angle of 59.35$\degree$. The band structure is computed from a non-self-consistent calculation along the two paths passing through high symmetry points listed in the Table~\ref{tab:tab_BZ_points}.

\begin{table}[h!]
\begin{center}
\begin{tabular}{c c c c || c c c c } 
 \hline
 & $\times$\textbf{b$_1$} & $\times$\textbf{b$_2$} & $\times$\textbf{b$_3$} & & $\times$\textbf{b$_1$} & $\times$\textbf{b$_2$} & $\times$\textbf{b$_3$} \\ 
 \hline\hline
 $P_1$ & -$\eta$ & $-\nu$ & $-\nu$ & $P_1$ & -$\eta$ & $-\nu$ & $-\nu$ \\ 
 $B_1$ & $-\eta$ & $-0.5$ & $\eta-1$ & $B$ & $-\eta$ & $\eta-1$ & $-0.5$ \\
 $F$ & $-0.5$ &  $-0.5$ & 0          & $F'$ & $-0.5$ &  0 & $-0.5$\\
 $B$ & $\eta-1$ & $-0.5$ & 1-$\eta$  & $B_1$ & $\eta-1$ & $1-\eta$ & $-0.5$\\
 $X$ &  0 & $-\nu$ & $\nu$           & $Q$ &  0 & $\nu$ & $-\nu$ \\
 $B$ & $1-\eta$ & $\eta-1$ & 0.5     & $B_1$ & $1-\eta$ & 0.5 & $\eta-1$ \\
 $F'$ & 0.5 & 0 & 0.5                & $F$ & 0.5 & 0.5 & 0 \\
 $B_1$ & $\eta$ & $1-\eta$ & 0.5     & $B$ & $\eta$ & 0.5 & $1-\eta$ \\
 $P$ & $\eta$ & $\nu$ & $\nu$        & $P$ & $\eta$ & $\nu$ & $\nu$ \\
\end{tabular}
\end{center}
\caption{Special points along the two considered paths at the boundary of the BZ of R3c BFO. The $\eta$ value is related to the rhombohedral angle $\alpha$ as $\eta=(1 + 4\cos(\alpha))/(2 + 4 \cos(\alpha))$ while $\nu=3/4-\eta/2$.}
\label{tab:tab_BZ_points}
\end{table}

\textit{Description of the sign inversion under the symmetry of altermagnetic BFO:} 
\setcounter{figure}{0}
\renewcommand{\figurename}{FIG. EM.}
\begin{figure*}[t]
\centering
\includegraphics[width=0.8\textwidth]{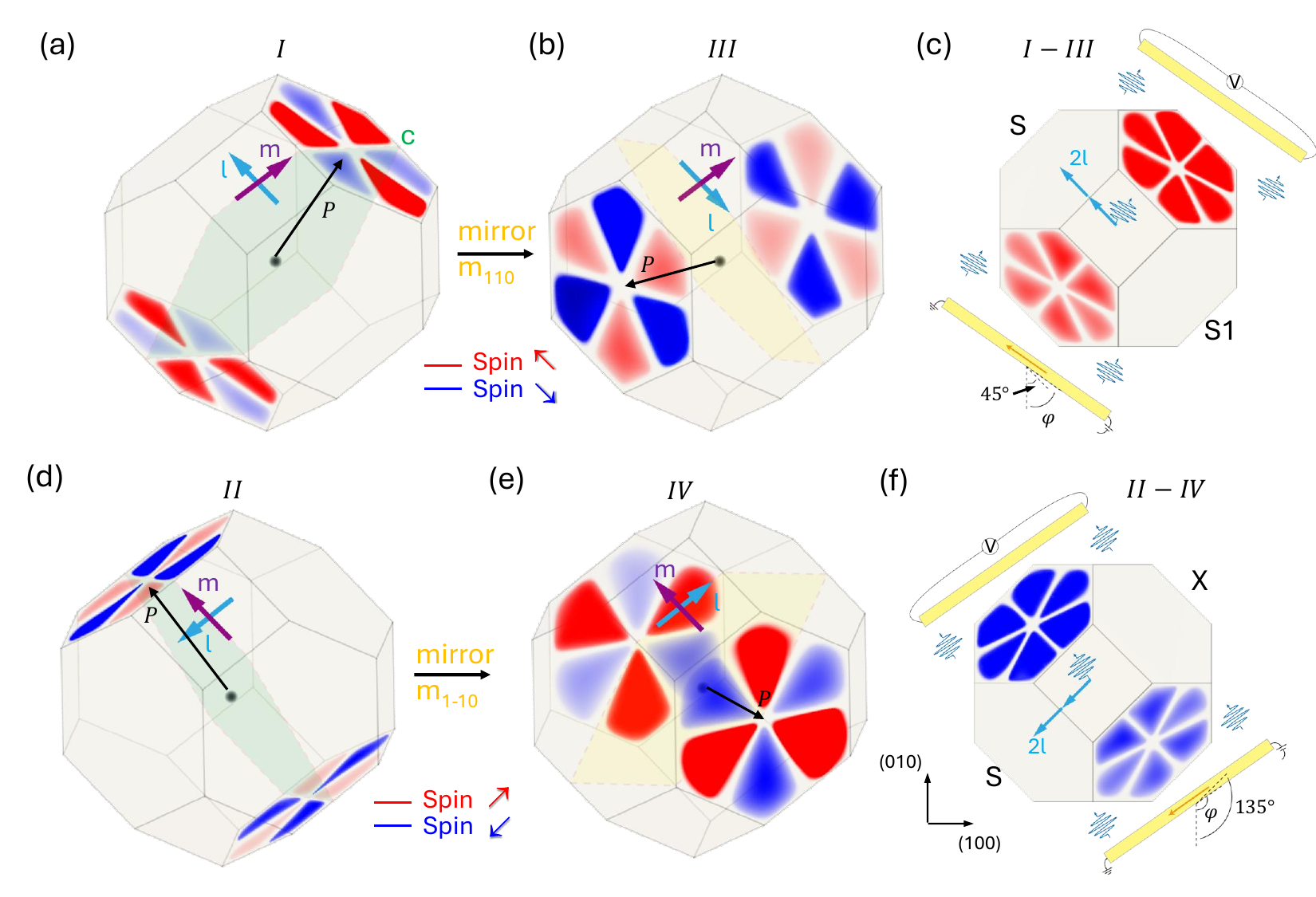}
\caption{\textbf{Possible altermagnetism contribution to the observed sign inversion in the spin transport.} Brillouin zone schematic of spin polarization corresponding to 45$\degree$ $I-III$ (a-c) and 135$\degree$ $II-IV$ (d-f). The green shaded planes in (a, d) represent the mirror-time-reversal symmetry of Cc' magnetic space group. When the ferroelectric polarization switches from I-III or II-IV, the magnetic orders are mirrored, represetned by yellow shaded planes in (b, e). The differential in (c) and (f) show the altermagnetism contribution to $\Delta V_{ISHE}$.}
\label{Fig:Fig.EM1}
\end{figure*}
We first consider whether the N\'{e}el vector and weak canted moment contribute to the sign inversion in the ISHE hysteresis shown in Fig. \ref{Fig:Fig.2}. The N\'{e}el vector \textit{l} is oriented perpendicular to the ferroelectric polarization, as illustrated in Fig. \ref{Fig:Fig.3}(a). This magnetic order belongs to the magnetic space group Cc' \cite{urruPRB_2025}, featuring a mirror-time-reversal glide plane, shown as green shaded plane in Fig \ref{Fig:Fig.3}(a). The time reversal symmetry is broken for this magnetic order since the two magnetic sublattices are surrounded by opposite tilt of oxygen octahedral sites, allowing for a weak canted moment induced by spin orbit coupling. The symmetry-allowed direction for the weak canted moment \textit{m} lies within the mirror–time-reversal glide plane, as shown in Fig \ref{Fig:Fig.3}(a) and (b). In our spin-transport measurements, the ISHE response arises from the spin projected onto the (100)/(010) BFO plane. Within this plane, the projected weak canted moment \textit{m} is parallel to the projected ferroelectric polarization and perpendicular to the projected N\'{e}el vector \textit{l}. Because the non-equilibrium magnon accumulation $\nabla(\mu_1 - \mu_2) \neq 0$, the two magnetic sublattices are not fully compensated during magnon transport. Therefore, we treat the N\'{e}el-vector contribution to magnon transport as an effective net magnetic moment.

The central question in interpreting the ISHE hysteresis is how the magnetic order switches when the ferroelectric polarization switches. As shown in the angular dependence of the ISHE hysteresis in Fig. \ref{Fig:Fig.2}(a), several angles exhibit zero crossings where $\Delta V_{ISHE}$ is negligible. For devices at these angles, the magnetic configuration either remains unchanged or becomes its mirror image when the ferroelectric polarization switches. These angles cluster near 45$\degree$, 90$\degree$, and 135$\degree$.

The sign reversal of $\Delta V_{ISHE}$ across the zero crossings near 45$\degree$ and 135$\degree$ indicates that, at these angles, the magnetic order switches to its mirror configuration. For example, at 45$\degree$, the ferroelectric polarization switches from configuration I to III. This ferroelectric polarization reversal can be viewed as a mirror operation across the (110) plane, as illustrated in Fig. EM.1(b), indicated by the yellow shaded plane. Under this mirror operation, the weak canted moment \textit{m}—being perpendicular to the mirror plane—remains unchanged, whereas the N\'{e}el vector \textit{l}, which lies within the plane, flips between configurations I and III. The same symmetry relation applies to the 135$\degree$ devices, in which magnetic order switches as a mirror operation between configurations II and IV.

For the 90$\degree$ device, the magnetic order switches between configurations II and III, or between configurations I and IV. Because $\Delta V_{ISHE}$ is minimum at this angle, these pairs of configurations must also be related by a mirror operation, since the N\'{e}el vector should keep perpendicular to the ferroelectric polarization \textit{P} and weak canted moment \textit{m}. The corresponding mirror operation is indicated by the black dashed line in Fig. \ref{Fig:Fig.3}(c). In our ISHE measurements, the Hall voltage is proportional to the magnitude of the magnetic moment projected along the spin-transport direction. For ferroelectric switching between configurations II and III, the projections of both the canted moment and the N\'{e}el vector are identical; the same holds for configurations I and IV. Thus, the mirror transformations enforce $\Delta V_{ISHE}$ minima for devices at 45$\degree$, 90$\degree$, and 135$\degree$.

We also highlight the small rhombohedral distortion, $\alpha \sim 0.6\degree$, shown in Fig. \ref{Fig:Fig.3}(c). This subtle distortion helps explain why the transformation from configuration I to II (a nearly 90\degree{} rotation about (001)) differs from that between II and III (a mirror operation), making the I-II and II-III switching inequivalent.

The magnetic configurations obtained through the mirror transformations are fully consistent, from a symmetry standpoint, with the observed sign reversal of $\Delta V_{ISHE}$ for devices oriented between 45$\degree$ and 135$\degree$. For devices with angles slightly larger than 45$\degree$, the switching still predominantly occurs between configurations I and III. In this case, the projected canted moment \textit{m} and the N\'{e}el vector \textit{l} are parallel to the spin-transport direction in configuration I (+\textit{P} state), but antiparallel in configuration III (-\textit{P} state). The resulting change in the magnitude of the spin projection produces a difference in the ISHE, leading to a hysteresis with a positive $\Delta V_{ISHE}$ during ferroelectric polarization switching. For devices with angles slightly smaller than 135$\degree$, the switching still mainly occurs between configurations II and IV. Here, the projected canted moment \textit{m} and N\'{e}el vector \textit{l} are antiparallel along the spin‐transport direction in configuration II (+\textit{P} state), but parallel in configuration IV (-\textit{P} state). This reversal of ferroelectric polarization, in contrast to the devices oriented slightly larger than 45$\degree$, now yields a negative $\Delta V_{ISHE}$ during the polarization switching.

The previous discussion focused on the static magnetic-moment contribution to the ISHE, which arises from the spin polarization at the $\Gamma$ point in the Brillouin zone. The R3c structure BFO with collinear magnetic order is expected to be a g-wave altermagnet. Fig. \ref{Fig:Fig.4}(d) presents the g-wave altermagnetic spin texture in the first Brillouin zone for magnetic configuration I. The g-wave altermagnetism is protected by the threefold rotational symmetry about the [111] axis and the three mirror planes intersecting the [111] direction, showing three nodal planes in the momentum space spin texture. The spin axes are along the N\'{e}el vector direction.

The sign reversal in $\Delta V_{ISHE}$ between configurations I-III and II-IV was attributed to the antiparallel orientations of the N\'{e}el vector \textit{l}. This N\'{e}el‐vector contribution further suggests that spin polarization away from the $\Gamma$ point in g-wave altermagnetism may also influence the observed ISHE, since the flowing magnon current shifts the momentum distribution away from $\Gamma$. Because the two magnetic sublattices are not fully compensated in magnon transport, we represent the resulting imbalance in spin population using strong and weak color contrasts in our Brillouin-zone illustrations.

For example, Fig. EM.1(a) shows a stronger red contrast than blue for configuration I, reflecting our assumption that spins parallel to the N\'{e}el vector contribute more strongly to spin transport and thus have a larger effective filling factor. In configuration III (Fig. EM.1(b)), the N\'{e}el vector flips, and consequently the blue contrast becomes stronger than the red. When magnons flow along the (110) direction in configuration I, the spins parallel to the N\'{e}el vector (red) are preferentially generated, whereas in configuration III the opposite spins (blue) dominate. Effectively, this magnon flow elongates the N\'{e}el vector, adding an extra contribution $\Delta l$.

This additional contribution, originating from momentum states away from the $\Gamma$ point, stems from the g-wave altermagnetic spin texture. For devices oriented slightly above 45$\degree$, both the N\'{e}el vector $l$ and the magnon-flow–induced $\Delta l$ contribute to $\Delta V_{ISHE}$. Subtracting configurations I and III yields the net difference in the magnon-induced $\Delta l$, as illustrated in Fig. EM.1(c). An analogous situation applies to devices slightly below 135$\degree$, which produces a contribution with the opposite sign after subtracting configurations II and IV (Fig. EM.1(f)), consistent with the reversed $\Delta V_{ISHE}$ dictates the altermagnetic nature of the underlying spin texture.

\end{document}